\documentclass[aps, pra, 10pt, twocolumn, superscriptaddress, longbibliography]{revtex4-1}
\usepackage{bm, color, amsmath, amssymb, graphicx}
\usepackage{umoline}

\newif\ifpng
\pngtrue
\begin{document}

\title{
	Crossing-Line-Node Semimetals:
	\\
	General Theory and Application to Rare-Earth Trihydrides
}

\author{Shingo Kobayashi}
\affiliation{Institute for Advanced Research, Nagoya University, Nagoya 464-8601, Japan}
\affiliation{Department of Applied Physics, Nagoya University, Nagoya 464-8603, Japan}
\author{Youichi Yamakawa}
\affiliation{Department of Physics, Nagoya University, Nagoya 464-8602, Japan}
\affiliation{Institute for Advanced Research, Nagoya University, Nagoya 464-8601, Japan}
\author{Ai Yamakage}
\affiliation{Department of Applied Physics, Nagoya University, Nagoya 464-8603, Japan}
\affiliation{Institute for Advanced Research, Nagoya University, Nagoya 464-8601, Japan}
\author{Takumi Inohara}
\affiliation{Department of Applied Physics, Nagoya University, Nagoya 464-8603, Japan}
\author{Yoshihiko Okamoto}
\affiliation{Department of Applied Physics, Nagoya University, Nagoya 464-8603, Japan}
\affiliation{Institute for Advanced Research, Nagoya University, Nagoya 464-8601, Japan}
\author{Yukio Tanaka}
\affiliation{Department of Applied Physics, Nagoya University, Nagoya 464-8603, Japan}

\begin{abstract}
	Multiple line nodes in energy-band gaps are found in semimetals preserving mirror-reflection symmetry. 
	We classify possible configurations of multiple line nodes with crossing points (crossing line nodes) under point-group symmetry. 
	Taking the spin-orbit interaction (SOI) into account, we also classify topological phase transitions from crossing-line-node Dirac semimetals to other topological phases, e.g., topological insulators and point-node semimetals.
	This study enables one to find crossing-line-node semimetal materials and their behavior in the presence of SOI from the band structure in the absence of SOI without detailed calculations.
	As an example, the theory applies to hexagonal rare-earth trihydrides with the HoD$_3$ structure and clarifies that it is a crossing-line-node Dirac semimetal hosting three line nodes. 
\end{abstract}

\maketitle

%\textbf{\textit{Introduction--.}}
\section{Introduction}
The degeneracy (node) of the energy spectrum in the Brillouin zone is a topological object. 
%$which is characterized by lower-dimensional topological invariants. 
Gapless semimetals are the realization of topological nodes in condensed matter physics \cite{murakami07, wan11, young12, fang12, wang12, chiu, chan16}.
Interestingly,
semimetals hosting topological nodes exhibit novel transport and response phenomena for external electromagnetic fields \cite{koshino10, zyuzin12, zyuzin12b, hosur13}.
For instance, in Weyl semimetals, which have point nodes in the Brillouin zone, electric current flows perpendicular to an electric field (anomalous Hall effect) and parallel to a magnetic field (chiral magnetic effect \cite{fukushima08}) due to the topological structure of the nodes.

Since topological invariants crucially depend on the spatial dimension \cite{schnyder08, kitaev09, ryu10, matsuura13}, node structures other than point nodes are expected to induce topological responses distinct from those in Weyl semimetals.
The line node \cite{burkov11, chiu14, fang15, gao16, fang16, yu16,carbotte16, lim17, roy, murakami} is one of these intriguing topological electronic states.
Many line-node semimetal materials 
\cite{
%mikitik06, heikkila11, carter12, heikkila,  
mullen15, chen, chen15, kim15, liu16, weng, huang16, xie15, yamakage16, ezawa16, zhu16, hirayama17, kawakami, xu17, guilehufe17, gupta} have been proposed and some measurements have actually seen line nodes in semimetals \cite{bian, schoop16, neupane16, wu16, hu, okamoto16, takane16, emmanouilidou}. 
Moreover, exotic magnetic transports \cite{singha, ali, wang, hu2} in line-node semimetals has been recently reported.
In addition, superconductivity is also found in the noncentrosymmetric line-node semimetal PbTaSe$_2$ \cite{wang16, zhang, chang, pang16, guan}.
Line-node semimetals have great potential for diverse developments in materials science.

In contrast to point nodes, there are many types of configurations of line nodes, i.e., single, spiral \cite{heikkila-kopnin-volovik11, heikkila-volovik11}, chain \cite{bzdusek}, separate multiple \cite{hirayama, bian, schoop16, neupane16, bian16}, nexus \cite{heikkila15, hyart, zhu}, and crossing \cite{weng15, zeng, kim-kane, yu15, du} line nodes.
%%%%%%%%%%%%%%%%%%%%%%%%%%%%%%%%%%%%%
\begin{figure}
\centering
\includegraphics[scale=0.5]{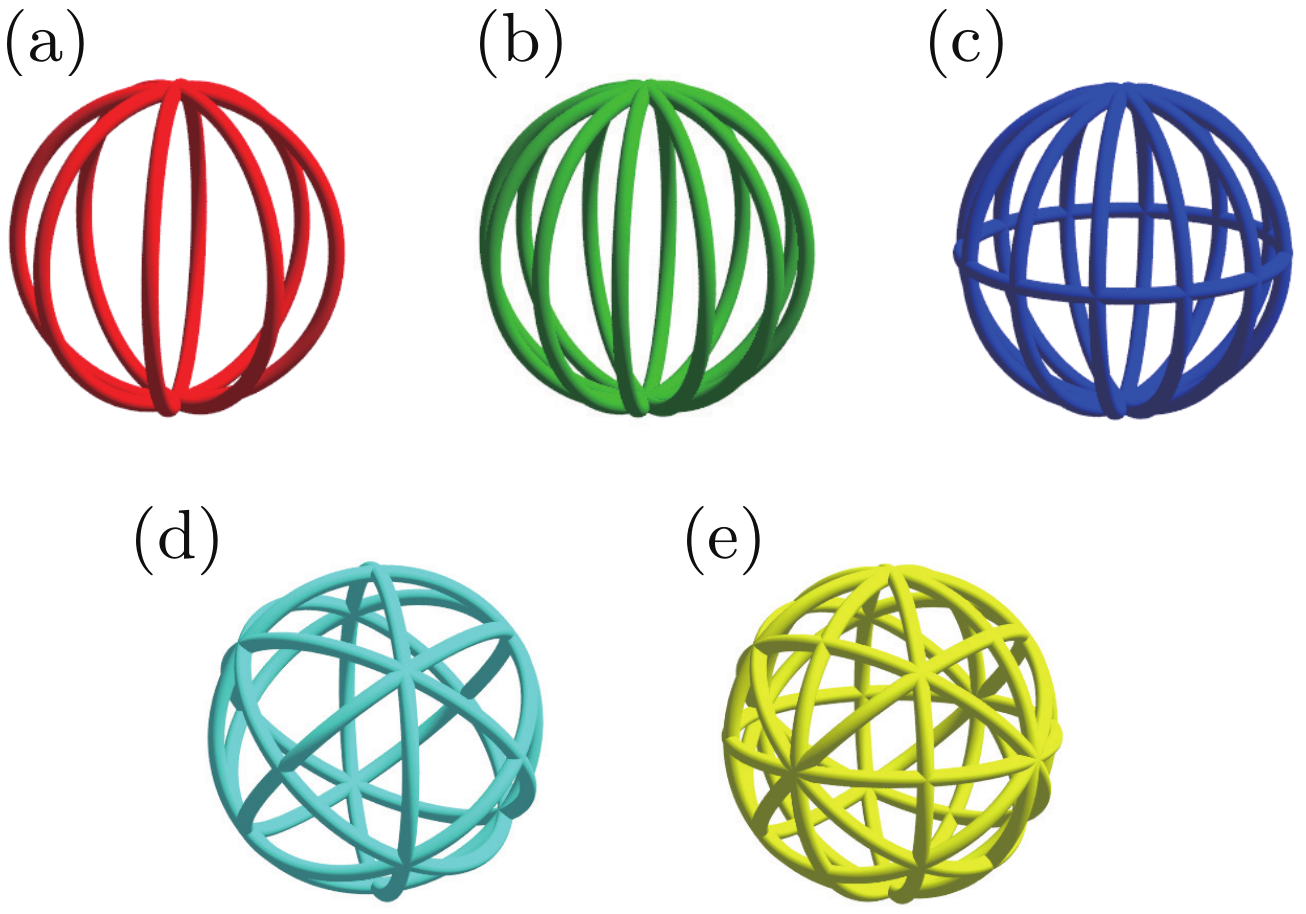}
\caption{Crossing line nodes in the momentum space.
(a)--(c) show three, six, and seven crossing line nodes realized in dihedral point-group symmetries.
(d) and (e) are six and nine crossing line nodes in cubic point-group symmetries.
}
\label{nodes}
\end{figure}
%%%%%%%%%%%%%%%%%%%%%%%%%%%%%%%%%%%%
In this work, we focus on crossing-line-node semimetals, as shown in Fig. \ref{nodes}, and study a general theory for it from the viewpoint of crystalline symmetry.
The configuration of the crossing line nodes is uniquely determined for a given level scheme of conduction and valence bands under a point-group symmetry.
The spin-orbit interaction (SOI) may open a gap in the line nodes but the crossing points possibly remain gapless, i.e., a Dirac semimetal may be realized.
We also clarify whether the resulting states are Dirac semimetals or (topological) insulators.
Applying the obtained results, one can find Dirac semimetals and topological insulators from line-node semimetals and can derive their topological indices from the band calculation in the \textit{absence} of SOI.

As an example, we apply the present theory to a hexagonal hydride, YH$_3$ [space group $P \bar 3 c 1$ (No. 165)], with the HoD$_3$ structure \cite{mansmann64}.
YH$_x$ has been focused on as a switchable mirror \cite{huiberts96}, i.e., the metal-insulator transition takes place at $x=2.85$ from a reflecting cubic crystal to a transparent hexagonal one. 
From optical measurements \cite{griessen97, gogh99, lee99, gogh01}, the gap has been evaluated to be 2.8 eV or slightly smaller.
On the other hand, early band calculations predicted that the hexagonal YH$_3$ is a semimetal rather than an insulator \cite{wang93, dekker93, wang95}.
Subsequent studies discussed another lower symmetric structure \cite{kelly97}, weak \cite{gelderen00, miyake00, gelderen02} and strong \cite{eder97, ng97, ng99} correlation effects giving rise to a finite gap in YH$_3$.
Although the actual material is insulating, 
we study the gapless electronic structure of the YH$_3$ without correlation effects, as a representative of HoD$_3$-structure materials, and its topological properties in detail since the electronic structure has been established so it is useful for further investigations.
The YH$_3$ with HoD$_3$ structure is shown to be a semimetal hosting three crossing line nodes.
A tiny energy gap ($\sim 4$ meV) is induced in the line nodes by SOI.
This gap is characterized by the topological indices of (1;000).

%\textbf{\textit{Crossing-line nodes protected by point group symmetries.--}}
\section{Crossing line nodes protected by point group symmetries}
%%%%%%%%%%%%%%
In general, a band crossing located on high-symmetry planes/lines is stable toward band repulsion if each energy band belongs to different eigenstates of crystalline symmetry. In particular, in mirror-reflection symmetric systems without SOI, a band crossing forms a stable Dirac line node (DLN) when it lies on a mirror-reflection plane and two energy bands have different mirror-reflection eigenvalues. 
 Generalizing this approach to all point groups, we investigate crossing line nodes protected by point groups: $C_{nv}$, $D_{nh}$, $D_{nd}$, $T_d$, $T_h$, and $O_h$ ($n=2,3,4,6$) and their possible topological phase transitions to topological insulators and Dirac semimetals.  
 
 Here, we consider a level scheme consisting of one-dimensional (1D) irreducible representations (IRRs) $\left( \Gamma_{1a},\Gamma_{1b} \right)$ of the lowest conduction and highest valence bands. 
 We focus on mirror-reflection symmetry-protected DLNs encircling time-reversal invariant momenta (TRIM). 
 According to the Schoenflies symbols, 
 mirror reflections are labeled as $\sigma_h$, $\sigma_v$, and $\sigma_d$, which represent horizontal, vertical, and diagonal mirror-reflection operations in point groups, respectively. 
 When conduction and valence bands cross on a $\sigma_{m}$ ($m=h,v,d$)-symmetric plane, the band crossing is stable if 1D IRRs $\Gamma_{1a}$ and $\Gamma_{1b}$ have different eigenvalues of $\sigma_m$ from each other, i.e., the character of $\sigma_m$ is $-1$ in $\Gamma_{1a}^{\ast} \times \Gamma_{1b}$. 
 Furthermore, the number of crossing lines corresponds to the number of equivalent $\sigma_m$ planes. 
 For example, in $C_{4v}$-symmetric systems, possible crossing-line-node configurations are $v^2$, $d^2$, and $v^2 d^2$ for $\{(A_1, B_2), (A_2, B_1)\}$, $\{(A_1, B_1), (A_2, B_2)\}$, and $\{(A_1, A_2), (B_1, B_2)\}$, respectively, where $v^i$ $(d^i)$ labels $i$ line nodes protected by $\sigma_v$ ($\sigma_d$) symmetry. Table~\ref{tab:PG-LN} shows possible crossing line nodes for each point group, and the correspondence with the level schemes is shown in {Appendix \ref{app1}}. 
 The symmetry-adapted effective Hamiltonian for 1D IRRs are also described in {Appendix \ref{app2}}. The study of crossing line nodes for 1D IRRs can be generalized to crossing line nodes for higher dimensional IRRs. In that case, it is necessary to take into account the effect of multibands. Nevertheless, when we choose a basis diagonalizing $\sigma_m$, the mechanism for protecting line nodes is the same as in the 1D IRR case: namely, a line node on a $\sigma_{m}$-symmetric plane is stable if two bands forming the line node have the different eigenvalues of $\sigma_m$. In particular, a level scheme consisting of 2D (3D) IRRs $\left( \Gamma_{2(3)a},\Gamma_{2(3)b} \right)$ leads to two (three) line nodes at most on a $\sigma_{m}$-symmetric plane. 
Possible line node configurations for 2D and 3D IRRs are listed in {Table \ref{tab:PG-higher} in Appendix.}
 
 \begin{table}[t]
 \centering
  \caption{Possible crossing line nodes and topological phase transitions for each point group (PG) for 1D IRRs. 
  In the second column, $m^i$ ($m=h$, $v$, $d$) indicates $i$ line nodes protected by $\sigma_m$. 
  The third column shows the effect of the SOI, where the DLNs encircle a TRIM. 
  TI and NI stand for the topological insulator and normal insulator, respectively. 
  DP stands for Dirac points, which are located on the $(n \geq 3)$-fold rotational axes.
  For the case of I, the SOI makes a gap on the crossing DLNs, but we cannot determine whether the system becomes a TI or NI from the point group symmetries.
  For the TI, the topological indices $(\nu_0; \nu_1 \nu_2 \nu_3)$ are obtained from Eqs. (\ref{nu0}) and (\ref{nui}).
  The configurations of $d^3$ ($v^3$), $d^3 v^3$, $h d^3 v^3$, $d^6$, and $h^3 d^6$ are depicted in Figs. \ref{nodes}(a)--\ref{nodes}(e), respectively.
  } \label{tab:PG-LN}
 \begin{tabular}{ccc}
  \hline\hline
  PG & Line nodes & SOI
  \\
  \hline
 $C_s$, $C_{nh}$ & $h$ & TI \\
 $D_{3d}$ & $d^3$ & TI\\
 $D_{nd}$ ($n=2$, $4$, $6$) & $d^n$ & I \\
 $C_{2v}$ & $v(zx)$, $v(zy)$ & TI \\
             & $v^2$ & I \\
 $C_{3v}$ & $v^3$ & TI \\
 $C_{nv}$ ($n=4$, $6$) & $v^{{n}/{2}}$, $d^{{n}/{2}}$& DP\\
             &  $v^{{n}/{2}} d^{{n}/{2}}$& I\\
 $D_{2h}$ & $h(zx)$, $h(zy)$, $h(xy)$, $h^3$ & TI \\
             & $h^2$ & NI \\
 $D_{3h}$ & $h$, $v^3$ & TI \\
            & $h v^3$ & I \\
 $D_{nh}$ ($n=4$, $6$) & $v^{{n}/{2}}$, $d^{{n}/{2}}$, $hv^{{n}/{2}}$, $hd^{{n}/{2}}$ & DP \\
             & $v^{{n}/{2}} d^{{n}/{2}}$ & NI \\
             & $h$, $h v^{{n}/{2}} d^{{n}/{2}}$ & TI \\
 $T_d$ & $d^6$& I \\
 $T_h$ & $h^3$& TI \\
 $O_h$ & $h^3$, $d^6$& DP\\
          & $h^3d^6$& TI \\
  \hline\hline
 \end{tabular}
\end{table}

\section{Effect of SOI}
In systems with SOI, mirror-reflection symmetry-protected line nodes are generally unstable \cite{yamakage16} except for nonsymmorphic systems \cite{fang15, liang16} since the mirror-reflection eigenvalues for spin up and down are different, i.e., $\Gamma_{1a}$ with spin up hybridizes with $\Gamma_{1b}$ with spin down.
This instability potentially leads to different topologically nontrivial phases such as Dirac/Weyl semimetals and topological insulators. 
The criteria for realizing these topological phases depend intrinsically on the level schemes and the number of line nodes encircling a TRIM, as we shall show in the following.

In the presence of SOI, the energy bands are labeled by the double representations, and 1D IRRs without SOI all become 2D IRRs after taking the product with the spin-$\frac{1}{2}$ representation $E_{1/2}$.
 Therefore, after including SOI, the crossing points of multiple line nodes on the $C_{nv}$-symmetric line remains as a Dirac point if each crossing energy band belongs to different double representations within $C_{nv}$, i.e., when $(\Gamma_{1a}',\Gamma_{1b}')$ in $C_{nv}$ are compatible with the 1D IRRs of $(\Gamma_{1a},\Gamma_{1b})$, and 
 $\Gamma_{1a}' \times E_{1/2}$ and $\Gamma_{1b}' \times E_{1/2}$ are different. Note that the $C_{nv}$-symmetry-protected Dirac points occur independently of the presence of spatial-inversion symmetry. The same criterion is applicable to higher dimensional IRRs if $\Gamma_{2(3) a} \times E_{1/2}$ is decomposed into 2D IRRs, and two different 2D IRRs cross on a $C_{nv}$-symmetric line. However, we do not completely predict the presence of Dirac points from the level schemes since the multibands are labeled again after including the SOI.  
Off the $C_{nv}$-symmetric line, antisymmetric SOI may turn line nodes into Weyl points.
The presence/absence of the Weyl points depends totally on the form of the SOI.
It is beyond the scope of the paper to discuss such Weyl points.
 
 % In particular, the doubly degenerate energy bands appear on the $C_n$-symmetric line when the system hosts $C_{nv}$ ($n=3,4,6$) or $C_i \times T$ due to the Kramer's degeneracy, where $T$ is the time-reversal operator. That is, after including the SOI, the crossing point of multiple line nodes remains as a Dirac point if $\Gamma_{1a} \times E_{1/2}$ and $\Gamma_{1b} \times E_{1/2}$ are different double representations in systems with $C_{nv}$ ($n=3,4,6$) or $C_i \times T$.
 
 If the SOI opens a gap on line nodes or an effect of breaking the crystalline symmetry destabilizes the Dirac point, the time-reversal-invariant systems potentially become topological insulators, depending on the band topology of the occupied states. For centrosymmetric systems with point groups $C_{n h}$, $D_{nh}$, $D_{3d}$, $T_h$, and $O_h$ ($n=2,4,6$), we can adapt the parity criterion proved in Ref.~\onlinecite{kim-kane} for the crossing line nodes, which allows us to determine the $\mathbb{Z}_2$ topological number $(\nu_0;\nu_1 \nu_2 \nu_3)$ of the topological insulator from the number of DLNs in the system without the SOI: (see {Appendix \ref{app3}} for more details)
 \begin{align}
  &\nu_0 = \sum_{n_1,n_2,n_3=0,1}N\left(\bm{\Gamma}_{(n_1 n_2 n_3)}\right) \ \ \mod 2, 
    \label{nu0}
  \\
   &\nu_i = \sum_{n_i =1;n_{j \neq i}=0,1} N\left(\bm{\Gamma}_{(n_1 n_2 n_3)}\right) \ \ \mod 2,
   \label{nui}
 \end{align}
 where $N(\bm{\Gamma}_{(n_1 n_2 n_3)})$ is the number of DLNs encircling the TRIM $\bm{\Gamma}_{(n_1 n_2 n_3)} =n_1 {\bm{b}_1}/{2} + n_2 {\bm{b}_2}/{2} + n_3 {\bm{b}_3}/{2}$ for $\bm{b}_i$ the $i$-th primitive reciprocal lattice vector. 
 
 On the other hand, for noncentrosymmetric systems, we can partially determine the $\mathbb{Z}_2$ topological numbers from the number of DLNs by adapting the mirror-parity criterion proved in Ref.\onlinecite{yamakage16}, which is applicable to the DLN $h$ of $C_s$, $C_{3h}$, and $D_{3h}$, $v$ of $C_{2v}$, and $v^3$ of $C_{3v}$ and $D_{3h}$. 
 %%%%%%%%%%%%%%%%%
 For these cases, the strong index $\nu_0$ is given by Eq. (\ref{nu0}).
 The weak indices $\nu_1$ and $\nu_2$ are given by Eq. (\ref{nui}).
 The third weak index $\nu_3$ is also determined from Eq. (\ref{nui}), except for $C_s$ and $C_{3h}$.
 %%%%%%%%%%%%%%%%%
 For example, when a single DLN encircles a TRIM $\bm{\Gamma}_{(n_1 n_2 n_3)}$ 
 %is located on the mirror-reflection plane of $n_3=0$ or $1$ 
 in the absence of SOI, 
 the $\mathbb{Z}_2$ topological numbers are given by ($1;n_1 n_2 \nu_{3}$) for $h$ of $C_s$ and $C_{3h}$; ($1;n_1n_2 n_3 $) for $h$ of $D_{3h}$,  $v$ of $C_{2v}$, and $v^3$ of $C_{3v}$ and $D_{3h}$, where $\nu_3$ is determined for $C_{2v}$, $C_{3v}$, and $D_{3h}$ due to the presence of an additional mirror-reflection symmetry. Other noncentrosymmetric systems are outside the scope of the mirror-parity criterion and depend on the details of the SOI.

\begin{table}
 \centering
 \caption{Proposed materials, configurations of crossing line nodes, resulting states induced by SOI, time-reversal-invariant momentum (TRIM) enclosed by the line nodes, and point group (PG) symmetry of the TRIM. 
 DP denotes the Dirac point. 
 MT carbon stands for Mackay--Terrones carbon.}
 \begin{tabular}{ccccccc}
  \hline\hline
  Material & LN & w/ SOI & TRIM & PG & Ref.
  \\
  \hline
  MT carbon & $h^3$ & DP & $R$ & $O_h$ & \onlinecite{weng15}
  \\
  LaN & $v^3$ & DP & $X$ & $D_{4h}$ & \onlinecite{zeng}
  \\
  Cu$_3$NPd & $h^3$ & DP & $R$ & $O_h$ & \onlinecite{kim-kane, yu15}
  \\
  CaTe & $hv^2$ & DP & $M$ & $D_{4h}$ & \onlinecite{du}
  \\
  YH$_3$ & $d^3$ & TI & $\Gamma$ & $D_{3d}$ & this work
  \\
  \hline\hline
 \end{tabular}
 \label{materials}
\end{table}
%%%%%%%%%%%%%%%%%%%%%%%%%%
%%%%%%%%%%%%%%%%%%%%%%%%%%
The obtained results enable us to predict the Dirac points and $\mathbb Z_2$ topological invariants in the presence of SOI from the band structures in the absence of SOI, without calculating the inversion/mirror-reflection parities of the wave functions.
As an example, in Table \ref{materials}, we show the results for four materials proposed in the literature.
    
%\textbf{\textit{One-dimensional $\mathbb Z_2$ invariant.--}}

%\textbf{\textit{Application to rare-earth trihydrides--.}}
\section{Application to rare-earth trihydrides}
Applying the general theory, we show that a hexagonal rare-earth trihydride with the HoD$_3$ structure is a crossing-line-node semimetal with three line nodes.
As a representative of the HoD$_3$-structure materials, we consider the hexagonal YH$_3$.
{Results for LuH$_3$ and ferromagnetic GdH$_3$ are shown in Appendix \ref{RH3}.}
In the present work, the band structure is calculated using the WIEN2k code \cite{WIEN2k}.
We used the full-potential linearized augmented plane-wave method within the generalized gradient approximation. 
{10 $\times$ 10 $\times$ 8} $k$ point sampling was used for the self-consistent calculation.

\begin{figure}
\centering
\includegraphics[scale=1.2]{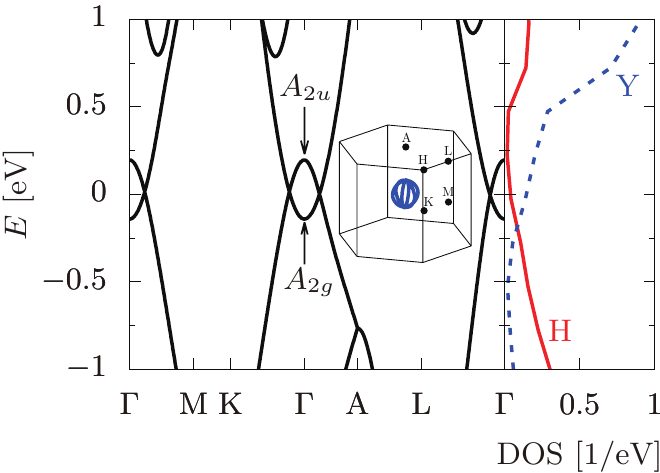}
\caption{Energy band and density of states of hexagonal YH$_3$. 
{The inset shows three crossing line nodes on the $M \Gamma A L$ planes, which corresponds to $d^3$ of $D_{3d}$ in Table \ref{tab:PG-LN} and Fig.~\ref{nodes}(a).}
The solid (red) and dashed (blue) lines denote the density of states of the H and Y atoms, respectively.
}
\label{dos}
\end{figure}

\begin{figure}
\centering
\includegraphics{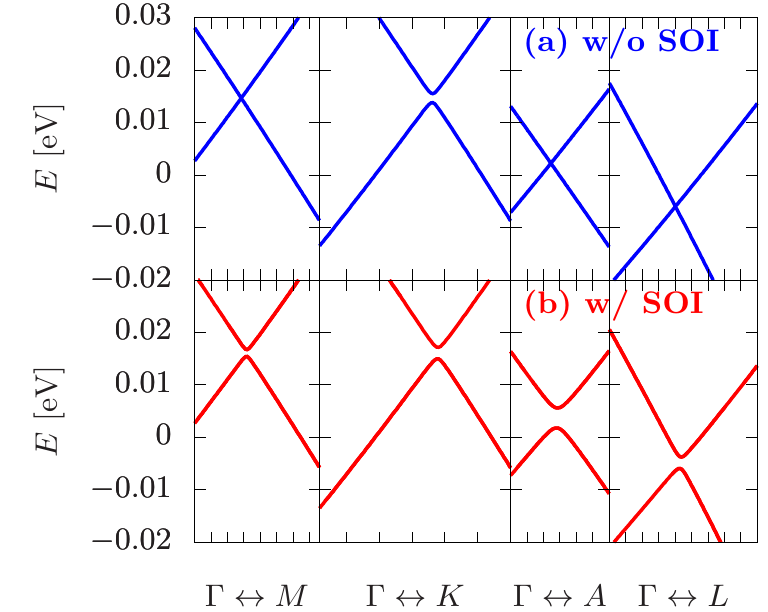}
\caption{Energy bands of YH$_3$ near the crossing line nodes (a) without and (b) with SOI.}
\label{band_detail}
\end{figure}

The gapless band structure in the hexagonal YH$_3$ was originally proposed by Dekker et al. \cite{dekker93} and is verified by our calculation, as shown in Fig. \ref{dos}.
%%%%%%%%%%%%%%%%%%%%%%%%%
Nearly gapless band dispersions are found on the $\Gamma$M, $\Gamma$K, and $\Gamma$A lines.
The detailed calculation shown in Fig. \ref{band_detail}(a) reveals that the band gap closes at 0.13 \AA$^{-1}$ on the $\Gamma M$ lines and at 0.14 \AA$^{-1}$ on the $\Gamma A$ lines while the gap opens by 4 meV on the $\Gamma$K line.
Moreover, the conduction and valence bands at the $\Gamma$ point are assigned to the $A_{2u}$ and $A_{2g}$ representations of $D_{3d}$, respectively.
%%%%%%%%%%%%%%%%%%%%%%%%%%%%%%%
From the general theory, the system must host three crossing line nodes in the A$_{2g}$--A$_{2u}$ scheme. 
Three crossing line nodes are actually seen on the three mirror (M$\Gamma$AL) planes. The location of the nodes is depicted in the inset of Fig. \ref{dos}.
On the $\Gamma$K line, a tiny band gap opens since the K$\Gamma$AH planes are not mirror planes. On other low-symmetry lines, the band gap is also weakly generated, on the order of 1 meV.
In other words, the system could behave as a Dirac-surface-node semimetal such as graphene networks \cite{zhong16} and Ba$MX_3$ ($M =$ V, Nb, Ta; $X =$ S, Se) \cite{liang16}, except for the low-energy and low-temperature regime (less than 1 meV).

It is worth mentioning that the Fermi surface of the hole-doped system mainly consists of the 1$s$ orbitals of the H atoms (see the right panel of Fig. \ref{dos}).
At $E_{\rm F} = -0.5$ eV, at which the carrier density is about $10^{20}$cm$^{-3}$, 90\% of the total density of states comes from the 1$s$ orbitals of H.
This Fermi surface might lead to high-temperature superconductivity, as in hydrogen sulfide \cite{drozdov15, akashi15, einaga16}.
Indeed, YH$_3$ has been predicted to be a superconductor below 40 K under 17.7 GPa \cite{kim09}, although the crystal structure is not the HoD$_3$ structure but the fcc under pressure \cite{ahuja97, palasyuk05, ohmura06, kume07, machida07}.

As mentioned above, the crossing line nodes realize and encircle the $\Gamma$ point, which has the $D_{3d}$-point-group symmetry.
The conduction and valence bands at the $\Gamma$ point are not degenerate, i.e., belong to the 1D IRRs of the $D_{3d}$ point group.
Then,
our general theory shown in Table \ref{tab:PG-LN} and Eqs. (\ref{nu0}) and (\ref{nui}) tells us that the SOI induces a gap on the crossing line nodes.
The resulting gapped state is a strong topological insulator of (1;000).
Notice that, strictly speaking, the system is semimetallic but the topological invariants are well defined since the direct gap opens at any momenta.
The first-principles data, which are shown in Fig. \ref{band_detail}(b), coincides with this prediction.
The induced spin-orbit gap is estimated to be on the order of 1 meV.
The SOI of the Y atom is small because it is not a heavy element.
The SOI of the H atom is, obviously, negligible.
Note that the Dirac point on the A point, which is located 0.7 eV below the Fermi level, still remains even in the presence of SOI, due to the nonsymmorphic symmetry of $P \bar 3 c 1$ \cite{young12}.

\begin{figure}
\centering
\ifpng
\includegraphics{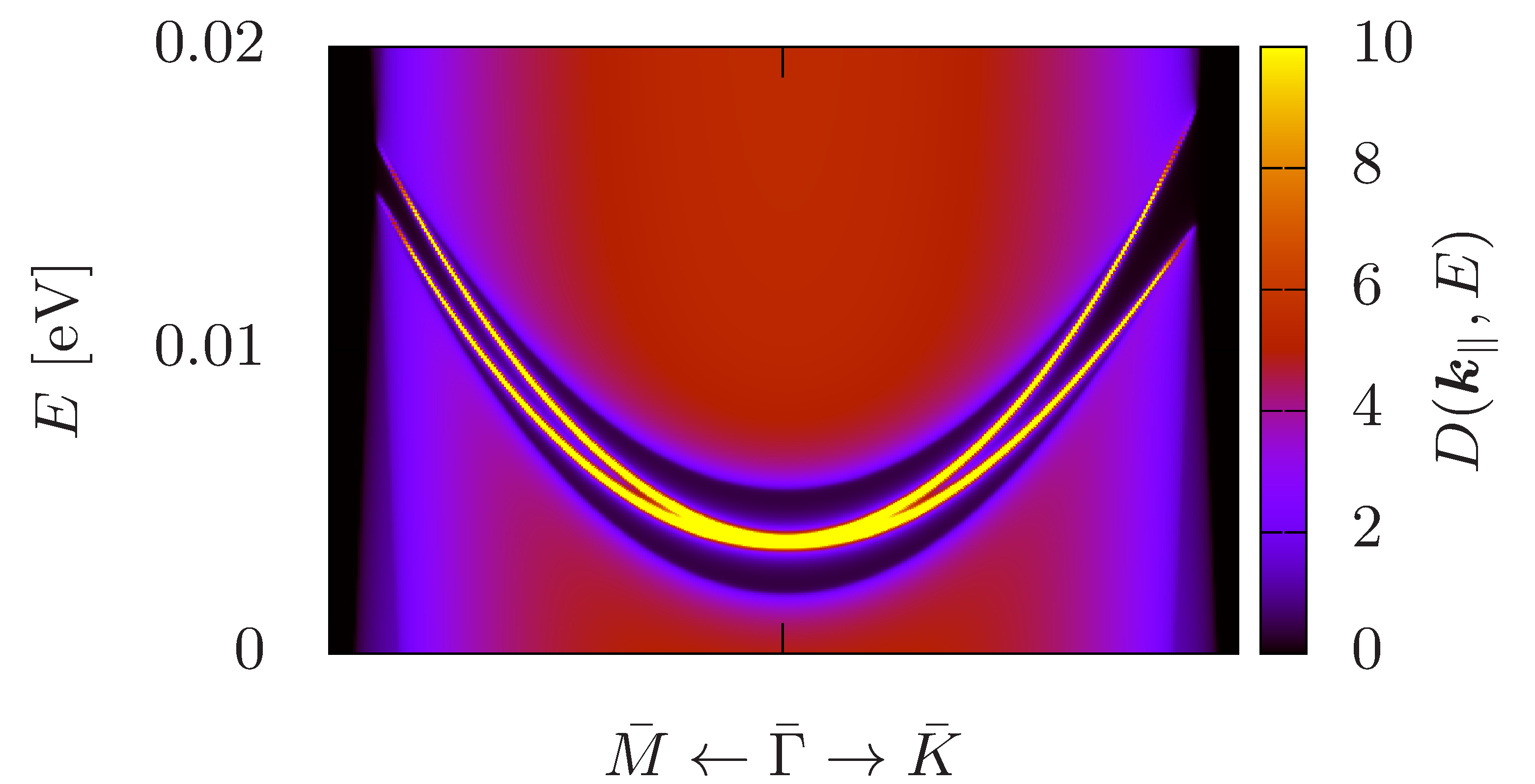}
\else
\includegraphics{yh3_surface.eps}
\fi
\caption{Angle-resolved density of states $D(\bm k_\parallel, E)$ on the (001) surface of YH$_3$. }
\label{yh3_surface}
\end{figure}

Finally, we construct a low-energy effective $k \cdot p$ Hamiltonian in the vicinity of the $\Gamma$ point to describe the crossing line nodes and SOI, as follows:
$
 H_0(\bm k)
% \nonumber\\  &
 = c(\bm k) \sigma_0 s_0 + m(\bm k) \sigma_3 s_0 + A (k_x^3-3k_x k_y^2) \sigma_2 s_0 + \mathcal O(k^4),
$ $
 H_{\rm SOI}(\bm k)
 =
 \lambda_1 \sigma_1 s_z k_z
 +
 \lambda_2 \sigma_1 (s_x k_x + s_y k_y) + \mathcal O(k^3),
$ 
$
 c(\bm k)
 = c_0 + c_1 k_z^2 + c_2 (k_x^2+k_y^2),
 \
 m(\bm k)
 =
 m_0 + m_1 k_z^2 + m_2 (k_x^2+k_y^2).
$
Here, $\sigma_i$ denotes the Pauli matrix for the orbitals ($\sigma_3 = \pm 1$ for the $A_{2g}$ and $A_{2u}$ orbitals, respectively).
$s_i$ denotes the Pauli matrix for the spin.
The parameters are determined to reproduce the crossing line nodes of the first-principles data:
$
 c_0 =  0.01391 \, \mathrm{eV},
 \
 c_1  = -0.5444 \, \mathrm{eV} {\mathrm{\mathring{A}^2}},
 \
 c_2 = 0.1185 \, \mathrm{eV} \mathrm{\mathring{A}}^2,
$
$
 m_0 = -0.15156 \, \mathrm{eV},
 \
 m_1 = 8.082 \, \mathrm{eV} \mathrm{\mathring{A}}^2,
 \
 m_2 = 8.314 \,\mathrm{eV} \mathrm{\mathring{A}}^2,
$
$A = 0.70$\,eV\AA$^3$,
 \
 $\lambda_1 = 0.01395$\,eV{\AA},
 \
 $\lambda_2 = 0.00621$\,eV{\AA}.
As seen in Fig. \ref{dos}, 
the band structure is nearly isotropic and particle-hole symmetric, hence the parameters approximately satisfy $m_1 \sim m_2$ and $c_i \ll m_i$.
Calculating the surface states of the above effective model, we verify that YH$_3$ is a strong topological insulator of $(1;000)$. 
We focus on the $(001)$ surface.
$k_z$ in the above Hamiltonian is regularized as $k_z \to  \sin(k_z c)/c$ and $k_z^2 \to 2[1-\cos(k_z c)]/c^2$.
The obtained lattice Hamiltonian is solved by using the recursive Green's function technique \cite{miyata13, miyata15}, and the angle-resolved density of states on the (001) surface is shown in Fig. \ref{yh3_surface}.
%%%%%%%%%%%%%%%%%%%%%%%%%%
%%%%%%%%%%%%%%%%%%%%%%%%%%
The system is, as mentioned above, a semimetal but hosts gapless surface states around the $\bar \Gamma$ point, which is projected from the $\Gamma$ point onto the surface, within the direct gap. 
This directly proves that the direct gap of YH$_3$ is characterized by the topological indices (1;000).

%\textbf{\textit{Summary--.}}
\section{Summary}
We studied a general theory classifying crossing-line-node semimetals under point-group symmetries.
The classification tells us the configuration of crossing line nodes for a given level scheme of conduction and valence bands.
This also enables us to determine whether the crossing line nodes are gapped by the SOI from the configuration of the nodes.
This will be quite important for materials development, i.e., one can predict materials being topological insulators and semimetals by exploring the band-calculation database in the absence of SOI, without any detailed calculations. 

We found that the rare-earth trihydride YH$_3$, as a representative of HoD$_3$--structure materials, is a crossing-line node semimetal, which hosts three line nodes on the mirror-reflection-invariant planes.
Although YH$_3$ is known to probably be an insulator by correlation effects, the present study encourages us to address materials with the HoD$_3$ structure and promises to realize a new topological semimetal.

This study has extensively revealed the electronic states of crossing line nodes.
There, on the other hand, remains an interesting issue: topological transports and responses in crossing-line-node semimetals.
The configuration is distinct from those of other point, line, and surface nodal structures.
Therefore, we expect new topological quantum phenomena in crossing-line-node semimetals, which should be clarified in future work.

\acknowledgments
This work was supported by the Grants-in-Aid for Young Scientists (B, Grant No. 16K17725),
 for Research Activity Start-up (Grant No. JP16H06861), and for Scientific Research on Innovative Areas ``Topological Material Science'' (JSPS KAKENHI Grants No. JP15H05851 and No. JP15H05853). S.K. was
 supported by the Building of Consortia for the Development of Human Resources in Science and Technology.

%\end{acknowledgment}

\appendix
% !TEX root = yh3170309.tex

\section{Tables of line node configurations for 1D IRRs}
\label{app1}

 Tables~\ref{tab:LS-Cnv}, \ref{tab:LS-Dnh}, and \ref{tab:LS-Oh} show the correspondence between the level schemes and line node configurations building on the criteria, where $m^i$ and $0$ indicate $i$ line nodes protected by $\sigma_m$ and the absence of a stable line node. After including the SOI, when crossing DLNs encircle a TRIM, they transform into normal (NI)/topological (TI) insulators or a Dirac point (DP). For the case of I, the SOI makes a gap on the crossing DLNs, but we cannot determine whether the system becomes a TI or NI from the point group symmetries. 
 
\begin{table}[h]
 \centering
  \caption{Line node configurations, level schemes, and the effect of SOI in $D_{nd}$ and $C_{nv}$ for 1D IRRs.} \label{tab:LS-Cnv} 
  \vspace{1ex}
     $D_{nd}$ ($n=2,4,6$)
    \\
  \begin{tabular}{c|cccc}
  \hline \hline 
        & $A_1$& $A_2$ &  $B_1$ & $B_2$  \\ \hline
   $A_1$&  0  & $d^n$/I & $d^n$/I & 0 \\  
   $A_2$&     &  0 & 0 & $d^n$/I \\
   $B_1$&    &  & 0 & $d^n$/I \\
   $B_2$&    &  &  & 0 \\
   \hline\hline
  \end{tabular}
 
 \vspace{1em}
 $D_{3d}$
 \\
  \begin{tabular}{c|cccc}
  \hline \hline
        & $A_{1g}$& $A_{2g}$ &  $A_{1u}$ & $A_{2u}$  \\ \hline
   $A_{1g}$&  0  & $d^3$/TI & $d^3$/TI & 0 \\  
   $A_{2g}$&     &  0 & 0 & $d^3$/TI \\
   $A_{1u}$&    &  & 0 & $d^3$/TI \\
   $A_{2u}$&    &  &  & 0 \\
   \hline\hline
  \end{tabular}
  
  \vspace{1em}
    $C_{2v}$
  \\
  \begin{tabular}{c|cccc}
    \hline\hline
    & $A_1$ & $A_2$ & $B_1$ & $B_2$
    \\
    \hline
    $A_1$ & 0 & $(xz)(yz)/\mathrm I$ & $(yz)/\mathrm I$ & $(xz)/\mathrm I$
    \\
    $A_2$ & & 0 & $(xz)/\mathrm I$ & $(yz)/\mathrm I$
    \\
    $B_1$ & & & 0 & $(xz)(yz)/\mathrm I$
    \\
    $B_2$ & & & & 0
    \\
    \hline\hline
  \end{tabular}

  \vspace{1em}
  $C_{3v}$
  \\
  \begin{tabular}{c|cc}
  \hline \hline
        & $A_{1}$& $A_{2}$ \\ \hline
   $A_{1}$&  0  & $v^3$/TI \\  
   $A_{2}$&     &  0 \\
   \hline\hline
  \end{tabular}
  
  \vspace{1em}
  $C_{2n v}$ ($n=2,3$)
  \\
  \begin{tabular}{c|cccc}
  \hline \hline
        & $A_{1}$& $A_{2}$ & $B_{1}$& $B_{2}$ \\ \hline
   $A_{1}$&  0  & $v^n d^n$/I & $d^n$/DP & $v^n$/DP \\  
   $A_{2}$&     &  0 & $v^n$/DP & $d^n$/DP \\
   $B_{1}$&     &     & 0 & $v^nd^n$/I \\
   $B_{2}$&     &  &  & 0 \\
   \hline\hline
  \end{tabular}
  \end{table}
 
 \begin{table*}[ht]
 \centering
  \caption{Line node configurations, level schemes, and the effect of SOI in $D_{nh}$ for 1D IRRs.} \label{tab:LS-Dnh}
  \vspace{1ex}
 $D_{2h}$
 \\
 \begin{tabular}{c|cccccccc}
  \hline \hline
        & $A_{g}$ & $B_{1g}$ &  $B_{2g}$ & $B_{3g}$ & $A_{u}$ & $B_{1u}$ &  $B_{2u}$ & $B_{3u}$ \\ \hline
   $A_{g}$  & 0 & $h^2$/NI & $h^2$/NI  & $h^2$/NI & $h^3$/TI & $h$/TI &  $h$/TI & $h$/TI \\  
   $B_{1g}$  &  & 0 &  $h^2$/NI & $h^2$/NI & $h$/TI & $h^3$/TI &  $h$/TI & $h$/TI \\
   $B_{2g}$ &  &  &  0 & $h^2$/NI & $h$/TI & $h$/TI &  $h^3$/TI & $h$/TI \\
   $B_{3g}$  & &  &   & 0 & $h$/TI & $h$/TI & $h$/TI & $h^3$/TI \\
   $A_{u}$  &  &  &   &  & 0 & $h^2$/NI &  $h^2$/NI & $h^2$/NI \\
  $B_{1u}$ &  &  &   &  &  & 0 &  $h^2$/NI & $h^2$/NI \\
   $B_{2u}$  &  &  &   &  &  &  &  0 & $h^2$/NI \\
   $B_{3u}$ &  &  &   &  &  &  &   & 0 \\
  \hline\hline
  \end{tabular}
  
  \vspace{1em}
  $D_{3h}$
  \\
  \begin{tabular}{c|cccc}
  \hline \hline
             & $A_{1}'$& $A_{2}'$ &  $A_{1}''$ & $A_{2}''$  \\ \hline
   $A_{1}'$&  0  & $v^3$/TI & $hv^3$/I & $h$/TI \\  
   $A_{2}'$&     &  0 & $h$/TI & $hv^3$/I \\
   $A_{1}''$&    &  & 0 & $v^3$/TI \\
   $A_{2}''$&    &  &  & 0 \\
   \hline\hline
  \end{tabular}
  
  \vspace{1em}
$D_{4h}$ 
\\
 \begin{tabular}{c|cccccccc}
  \hline \hline
        & $A_{1g}$ & $A_{2g}$ &  $B_{1g}$ & $B_{2g}$ & $A_{1u}$ & $A_{2u}$ &  $B_{1u}$ & $B_{2u}$\\ \hline
   $A_{1g}$  & 0 & $v^2 d^2$/NI & $d^2$/DP & $v^2$/DP & $hv^2d^2$/TI & $h$/TI &  $hv^2$/DP & $hd^2$/DP \\  
   $A_{2g}$  &  & 0 &  $v^2$/DP & $d^2$/DP & $h$/TI & $v^2d^2$/NI &  $hd^2$/DP & $hv^2$/DP \\
   $B_{1g}$ &  &  &  0 & $v^2d^2$/NI & $hv^2$/DP & $hd^2$/DP &  $hv^2d^2$/TI & $h$/TI \\
   $B_{2g}$ & &  &   & 0 & $hd^2$/DP & $hv^2$/DP &  $h$/TI & $hv^2d^2$/TI \\
   $A_{1u}$  &  &  &   &  & 0 & $v^2d^2$/NI &  $d^2$/DP & $v^2$/DP \\
   $A_{2u}$ &  &  &   &  &  & 0 &  $v^2$/DP & $d^2$/DP \\
   $B_{1u}$  &  &  &   &  &  &  &  0 & $v^2d^2$/NI \\
   $B_{2u}$  &  &  &   &  &  &  &   & 0 \\
  \hline\hline
  \end{tabular}
  
  \vspace{1em}
  $D_{6h}$ 
  \\
 \begin{tabular}{c|cccccccc}
  \hline \hline
        & $A_{1g}$ & $A_{2g}$ &  $B_{1g}$ & $B_{2g}$ & $A_{1u}$ & $A_{2u}$ &  $B_{1u}$ & $B_{2u}$\\ \hline
   $A_{1g}$  & 0 & $v^3 d^3$/NI & $hv^3$/DP & $hd^3$/DP & $hv^3d^3$/TI & $h$/TI &  $d^3$/DP & $v^3$/DP \\  
   $A_{2g}$  &  & 0 &  $hd^3$/DP & $hv^3$/DP & $h$/TI & $hv^3d^3$/TI &  $v^3$/DP & $d^3$/DP \\
   $B_{1g}$ &  &  &  0 & $v^3d^3$/NI & $d^3$/DP & $v^3$/DP &  $hv^3d^3$/TI & $h$/TI \\
   $B_{2g}$ & &  &   & 0 & $v^3$/DP & $d^3$/DP &  $h$/TI & $hv^3d^3$/TI \\
   $A_{1u}$  &  &  &   &  & 0 & $v^3 d^3$/NI &  $hv^3$/DP & $hd^3$/DP \\
   $A_{2u}$ &  &  &   &  &  & 0 &  $hd^3$/DP & $hv^3$/DP \\
   $B_{1u}$  &  &  &   &  &  &  &  0 & $v^3d^3$/NI \\
   $B_{2u}$  &  &  &   &  &  &  &   & 0 \\
  \hline\hline
  \end{tabular}
  \end{table*}
  
  \begin{table}
 \centering
  \caption{Line node configurations, level schemes, and the effect of SOI in $T_d$, $T_h$, and $O_h$ for 1D IRRs.} \label{tab:LS-Oh}
    $T_d$
    \\
  \begin{tabular}{c|cc}
  \hline \hline
        & $A_{1}$& $A_{2}$  \\ \hline
   $A_{1}$&  0  & $d^6$/I \\  
   $A_{2}$&     &  0 \\
   \hline\hline
  \end{tabular}
  
  \vspace{1em}
     $T_h$
     \\
  \begin{tabular}{c|cc}
  \hline \hline
        & $A_{g}$& $A_{u}$ \\ \hline
   $A_{g}$&  0  & $h^3$/TI \\  
   $A_{u}$&     &  0 \\
   \hline\hline
  \end{tabular}
  
  \vspace{1em}
   $O_h$
   \\
  \begin{tabular}{c|cccc}
  \hline \hline
        & $A_{1g}$& $A_{2g}$ &  $A_{1u}$ & $A_{2u}$  \\ \hline
   $A_{1g}$&  0  & $d^6$/DP & $h^3d^6$/TI & $h^3$/DP \\  
   $A_{2g}$&     &  0 & $h^3$/DP & $h^3d^6$/TI \\
   $A_{1u}$&    &  & 0 & $d^6$/DP \\
   $A_{2u}$&    &  &  & 0 \\
   \hline\hline
  \end{tabular}
 
\end{table}

\section{Symmetry-adapted effective models}
\label{app2}

\begin{table*}[t]
 \centering
  \caption{Symmetry-adapted $f_+ (\bm{k})$ for each line node configuration in point groups (PGs), where $k_{\pm} = k_x \pm ik_y$. 
  We show $f_+(\bm k)$ for $d^n$ of $D_{nd}$ when $U(C_2')$ is given by
  $U(C_2') = \pm \sigma_0$.
  } 
 \label{tab:PG-model}
 \begin{tabular}{ccc}
  \hline\hline
  PG & Line nodes & $f_+ (\bm{k})$
  \\
  \hline
 $C_s$,$C_{nh}$ & $h$ & $k_z$ \\
 $D_{3d}$ & $d^3$ & ${\rm Re} \left[ k_+^n \right]$ \\
 $D_{nd}$ ($n=2,4,6$) & $d^n$ & ${\rm Im} \left[ k_+^n \right]$ \\
 $C_{2v}$ & $(xz)$; $(yz)$; $(yz)(xz)$ & $k_y$; $k_x$; $k_xk_y$ \\
 $C_{3v}$ & $v^3$ & ${\rm Re} \left[k_+^3\right] $ \\
 $C_{4v}$ & $v^{2}$; $d^{2}$; $v^{2} d^{2}$& ${\rm Im} \left[k_+^2\right]$; ${\rm Re} \left[k_+^2\right]$; ${\rm Re} \left[k_+^2\right] {\rm Im} \left[k_+^2\right]$\\
 $C_{6v}$ & $v^{3}$; $d^{3}$; $v^{3} d^{3}$& ${\rm Re} \left[k_+^3\right]$; ${\rm Im} \left[k_+^3\right]$; ${\rm Re} \left[k_+^3\right] {\rm Im} \left[k_+^3\right]$\\
  $D_{2h}$ & $(xz)$; $(yz)$; $(xy)$;  
  & $k_y$; $k_x$; $k_z$;  \\
 & $(xz)(yz)$; $(yz)(xy)$; $(xz)(xy)$; $(yz)(xz)(xy)$
 &
 $k_y k_x$; $k_x k_z$; $k_y k_z$; $k_xk_yk_z$
  \\
 $D_{3h}$ & $h$; $v^3$; $h v^3$ & $k_z$; ${\rm Re} \left[k_+^3\right] $; $k_z{\rm Re} \left[k_+^3\right] $  \\
 $D_{4h}$ & $h$; $v^{2}$; $d^{2}$;  $v^{2} d^{2}$;
 & 
 $k_z$; ${\rm Im} \left[k_+^2\right]$; ${\rm Re} \left[k_+^2\right]$; ${\rm Re} \left[k_+^2\right] {\rm Im} \left[k_+^2\right]$;
  \\
 &
  $hv^{2}$; $hd^{2}$; $hv^{2} d^{2}$
 &
  $k_z{\rm Im} \left[k_+^2\right]$;
  $k_z{\rm Re} \left[k_+^2\right]$; $k_z{\rm Re} \left[k_+^2\right] {\rm Im} \left[k_+^2\right]$
 \\
 $D_{6h}$ & $h$; $v^{3}$; $d^{3}$; 
 & 
 $k_z$; ${\rm Re} \left[k_+^3\right]$; ${\rm Im} \left[k_+^3\right]$; 
 \\
 & $v^{3} d^{3}$; $hv^{3}$; $hd^{3}$; $hv^{3} d^{3}$
 &
 ${\rm Re} \left[k_+^3\right] {\rm Im} \left[k_+^3\right]$; $k_z{\rm Re} \left[k_+^3\right]$; $k_z{\rm Im} \left[k_+^3\right]$; $k_z{\rm Re} \left[k_+^3\right] {\rm Im} \left[k_+^3\right]$
 \\
 $T_d$ & $d^6$& $\left(k_x^2-k_y^2\right)\left(k_y^2-k_z^2\right)\left(k_z^2-k_x^2\right)$
 \\
 $T_h$ & $h^3$& $k_xk_yk_z$ \\
 $O_h$ & $h^3$; $d^6$;  & $k_xk_yk_z$; $\left(k_x^2-k_y^2\right)\left(k_y^2-k_z^2\right)\left(k_z^2-k_x^2\right)$; 
 \\
 & $h^3d^6$
 &
 $k_x k_y k_z\left(k_x^2-k_y^2\right)\left(k_y^2-k_z^2\right)\left(k_z^2-k_x^2\right) $
 \\
  \hline\hline
 \end{tabular}
\end{table*}

First of all, consider a level scheme consisting of 1D IRRs $\Gamma_{1a}$ and $\Gamma_{1b}$. The low-energy effective Hamiltonian is generally described by
\begin{align}
 H(\bm{k}) = f_0 (\bm{k}) \sigma_0 + f_z (\bm{k}) \sigma_z + f_+(\bm{k}) \sigma_+ + f_+(\bm{k})^{\ast} \sigma_-, \label{eq:A-Hami}
\end{align}
where $(\sigma_0,\bm{\sigma})$ are the $2\times 2$ identity and Pauli matrices in the orbital space and $\sigma_{\pm} = (\sigma_x \pm i \sigma_y)/2 $. We assume that the Hamiltonian (\ref{eq:A-Hami}) possesses time-reversal symmetry, which demands that $f_0(\bm{k})^{\ast}=f_0(-\bm{k})$, $f_z (\bm{k})^{\ast}=f_z (-\bm{k})$, and $f_+ (\bm{k})^{\ast}=f_+ (-\bm{k})$. The group operation on this Hamiltonian is defined by
\begin{align}
 U(g)^{\dagger} H(\bm{k})U(g) = H(D(g) \bm{k}), \label{eq:A-go}
\end{align}
where $U(g)$ is a unitary matrix in terms of $g$ in the orbital space and $D(g)$ represents a rotation matrix concerning $g$ in the momentum space. Since we focus on the 1D IRRs, $U(g)$ becomes $\pm \sigma_0$ or $\pm \sigma_z$. In particular, the mirror-reflection operations $\sigma_h$, $\sigma_v$, and $\sigma_d$ are given as follows:
\begin{itemize}
\item $\sigma_h$ in $C_s$, $C_{nh}$, $D_{nh}$, $T_h$, and $O_h$:
\begin{align}
&
 U(\sigma_h)^{\dagger} H(k_x,k_y,k_z) U(\sigma_h) 
 \nonumber\\ & \hspace{8em}
 = H(k_x,k_y,-k_z).
\end{align}
\item $\sigma_v(yz)$ in $C_{2v}$ and $D_{2h}$; $\sigma_v$ in $C_{nv}$ and $D_{nh}$ ($n=3,4,6$); $\sigma_d$ in $D_{ld}$ ($l=2,3,4,6$):
\begin{align}
&
 U(\sigma_{v(d)})^{\dagger} H(k_x,k_y,k_z) U(\sigma_{v(d)}) 
 \nonumber\\ & \hspace{8em}
 = H(-k_x,k_y,k_z).
\end{align}
\item $\sigma_v(xz)$ in $C_{2v}$ and $D_{2h}$; $\sigma_d$ in $C_{6v}$ and $D_{6h}$:
\begin{align}
&
 U(\sigma_{v(d)})^{\dagger} H(k_x,k_y,k_z) U(\sigma_{v(d)}) 
 \nonumber\\ & \hspace{8em}
 = H(k_x,-k_y,k_z).
\end{align}
\item $\sigma_d$ in $C_{4v}$, $D_{4h}$, $T_d$, and $O_h$:
\begin{align}
&
 U(\sigma_d)^{\dagger} H(k_x,k_y,k_z) U(\sigma_d) 
 \nonumber\\ & \hspace{8em}
 = H(k_y,k_x,k_z).
\end{align}
\end{itemize}
 Assuming that the crossing energy bands appear around the $\sigma_m$ ($m=h,v,d$)-symmetric planes, the crossing line is stable if $f_+=0$ on the $\sigma_m$-symmetric planes because the $f_+$ term describes the band mixing between $\Gamma_{1a}$ and $\Gamma_{1b}$ and makes a gap. Thus, the stable DLNs requires that $U(\sigma_m) = \pm \sigma_z$, leading to $f_+(\bm{k}_{\parallel},k_{\perp}) = -f_+(\bm{k}_{\parallel},-k_{\perp}) $, where $\bm{k}_{\parallel}$ and $k_{\perp}$ are the momenta parallel to and perpendicular to the $\sigma_m$-symmetric planes. This condition is consistent with the criterion in the main paragraph. Table~\ref{tab:PG-model} shows the symmetry-adapted $f_+ (\bm{k})$ for each line node configuration.

\begin{figure}
\centering
\includegraphics[width=8cm]{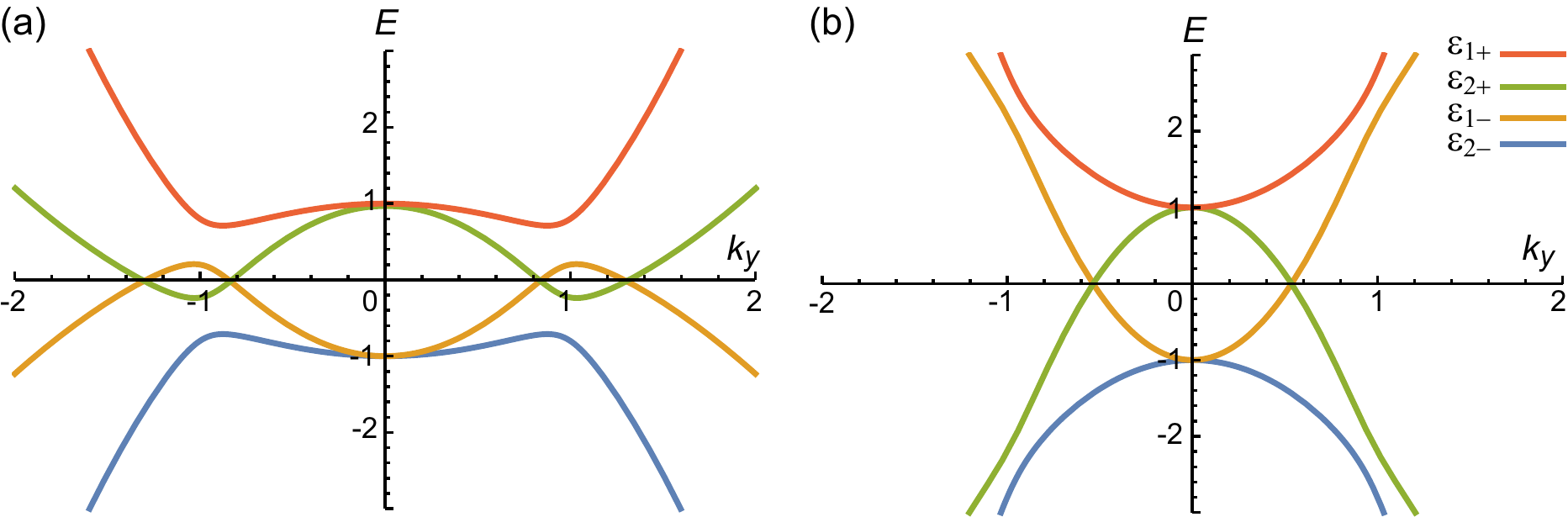}
\caption{
Evolution from double line nodes $2 v^3$ to single line node $v^3$.
The energy bands of the effective model for 2D IRRs defined by Eqs. (\ref{eq:2DIRR_e1}) and (\ref{eq:2DIRR_e2}) with parameters $(c_0,c_2,c_0',c_2',k_z) = (-1,1,1,-1,0)$. $v_1=0.5$ and $v_2 = 0.3$ for (a). $v_1=2.5$ and $v_2 = 0.3$ for (b).  }
\label{fig:band_2DIRRs}
\end{figure}

 Next, consider a level scheme consisting of 2D (3D) IRRs ($\Gamma_{2a},\Gamma_{2b}$) [($\Gamma_{3a},\Gamma_{3b}$)]. To avoid cumbersome multiband effects, we ignore the level splitting and consider doubly (triply) degenerate conduction and valence bands as a starting point. In that case, the energy bands all form a DLN, and it is possible to decompose the effective Hamiltonian for 2D (3D) IRRs into two (three) $2 \times 2$ effective Hamiltonians in terms of 1D IRRs. As an example of 2D IRRs, we discuss the level scheme ($E,E$) for $C_{3v}$. The symmetry operators are defined as
 \begin{align}
  &U(\sigma_d) = {\rm diag}[1,-1,1,-1], \\
   &U(C_3) =\frac{1}{2}\begin{pmatrix} -1&-\sqrt{3}&0&0 \\ \sqrt{3} & -1 &0&0 \\ 0&0&-1&-\sqrt{3} \\ 0&0&\sqrt{3}&-1 \end{pmatrix}.
 \end{align}
Then, the symmetry-adapted effective Hamiltonian is given by
 \begin{align}
  H(\bm{k}) = \begin{pmatrix} f_{0}(\bm{k}) &0&0
  &
  i v {\mathrm{Re}} [k_+^3] \\ 0 & f_{0}(\bm{k}) 
  &
  - i v{\mathrm{Re}} [k_+^3]&0  \\ 0
  &
  iv \mathrm{Re} [k_+^3] & g_{0}(\bm{k}) &0   
  \\ -iv \mathrm{Re} [k_+^3]&0&0& g_{0}(\bm{k})  \end{pmatrix},
 \end{align} 
 where $f_0(\bm{k})=c_0+c_1 k_z^2 +c_2 (k_x^2+k_y^2)$ and $g_0(\bm{k})=c_0'+c_1' k_z^2 +c_2' (k_x^2+k_y^2)$. Here, $c_0$, $c_1$, $c_2$, $c_0'$, $c_1'$, $c_2'$, and $v$ are material dependent parameters.
The effective Hamiltonian can be described by the block-diagonal form: $H_{+v^3}(\bm{k}) \oplus H_{-v^3}(\bm{k})$, where $H_{\pm v^3}(\bm{k})$ is a $2 \times 2$ effective Hamiltonian with $f_{+} (\bm{k}) = \pm i v {\rm Re} [k_+^3]$. When $f_0$ and $g_0$ cross on the $\sigma_v$-symmetric planes, we obtain six DLNs and label this line node configuration as $2 v^3$, where $n m^i$ represents $i$ $n$th-degenerate DLNs protected by $\sigma_m$, i.e., $n \times i$ line nodes appear on the $\sigma_m$-symmetric plane. To check the effect of band splitting, we include it as a perturbation: $H(\bm{k})+H'(\bm{k})$ with
 \begin{align}
 &
 H'(\bm{k}) 
 \nonumber\\ &
 =  \begin{pmatrix} v_1 {\rm Re}[k_+^2] &  v_1 {\rm Im}[k_-^2] & v_2 {\rm Re}[k_+^2]  & v_2 {\rm Im}[k_-^2] \\
  v_1 {\rm Im}[k_-^2] & -v_1  {\rm Re}[k_+^2]  & v_2 {\rm Im}[k_-^2] & -v_2 {\rm Re}[k_+^2]  \\
   v_2 {\rm Re}[k_+^2]  & v_2 {\rm Im}[k_-^2] &  v_1 {\rm Re}[k_+^2]  & v_1  {\rm Im}[k_-^2]  \\
    v_2 {\rm Im}[k_-^2] & -v_2 {\rm Re}[k_+^2]  &  v_1  {\rm Im}[k_-^2]  & - v_1 {\rm Re}[k_+^2]  \end{pmatrix}.
 \end{align}
 Since the three $\sigma_v$-symmetric planes are equivalent, we focus on the $\sigma_v$-symmetric plane of $k_x=0$, on which the eigenvalues of $H(\bm{k})+H'(\bm{k})$ are
 \begin{align}
  &\epsilon_{1\pm}(\bm{k}) = \frac{f_0+g_0}{2} + v_1 k_y^2 \pm \sqrt{\frac{(f_0-g_0)^2}{4} + v_2^2 k_y^2}, \label{eq:2DIRR_e1}\\
  &\epsilon_{2\pm}(\bm{k}) = \frac{f_0+g_0}{2} - v_1 k_y^2 \pm \sqrt{\frac{(f_0-g_0)^2}{4} + v_2^2 k_y^2}. \label{eq:2DIRR_e2}
 \end{align} 
 The energy bands are plotted in Fig.~\ref{fig:band_2DIRRs}. The small band splitting does not break the $2 v^3$ line node structure when $v_1 > v_2$ 
 [see Fig.~\ref{fig:band_2DIRRs} (a)]. 
 On the other hand, for large $v_1$, $2 v^3$ changes to $v^3$ due to the change in band structure 
 [see Fig.~\ref{fig:band_2DIRRs} (b)]. Thus, although there exist at most two line nodes on a $\sigma_v$-symmetric plane, we can engineer the line node configuration from $2v^3$ to $v^3$ or $0$ by the band splitting $H'(\bm k)$. 
 In a similar manner, we can construct symmetry-adapted effective models for 3D IRRs. For example, consider the level scheme consisting of ($T_{2g},T_{1u}$) of $O_h$. In this case, the Hamiltonian is block-diagonalized as $H_{h^3}(\bm{k}) \oplus H_{h^3}(\bm{k}) \oplus H_{h^3}(\bm{k})$, where $H_{h^3}(\bm{k})$ is a $2 \times 2$ effective Hamiltonian with $f_+(\bm{k}) = v k_x k_y k_z$. Thus, we obtain nine DLNs, labeled by $3 h^3$. After including the effect of band splitting, $3h^3$ changes to $2 h^3$, $h^3$, or $0$.  In general, the decomposition of level schemes ($\Gamma_{2(3)a},\Gamma_{2(3)b}$) into $H_{m^i} (\bm{k})$ is possible if  $\Gamma_{2(3)a}^{\ast} \times \Gamma_{2(3)b}$ includes a 1D IRR whose character of $m^i$ is $-1$. We list possible decompositions for level schemes with 2D and 3D IRRs in Table~\ref{tab:PG-higher}. Our method derives symmetry-adapted effective models in a comprehensive fashion, but accidental line nodes often occur off mirror-reflection symmetric planes. 
 
  \begin{table*}[t]
 \centering
  \caption{Possible line node configurations for level scheme with 2D and 3D IRRs, where we ignore the effect of band splittings. $n m^i$ represents $i$ $n$th-degenerate line nodes protected by $\sigma_m$ on $\sigma_m$-symmetric planes, i.e., $n \times i$ line nodes appear in total. Taking into account band splittings, the line node configurations change from $n m^i $ to $(n-1) m^i$, $(n-2) m^i$, $\cdots$, $ m^i$, or $0$, depending on the magnitude of the band splittings.  } \label{tab:PG-higher}
 \begin{tabular}{ccc}
  \hline\hline
  PG & Level scheme & Line nodes
  \\
  \hline
  $C_{3h}$ & ($E',E''$) & $2h$ \\
  $C_{4h}$ & $\left(E_g,E_u\right)$  & $2 h$ \\
  $C_{6h}$ & $\left(E_{1g(u)},E_{2g(u)} \right)$, $\left(E_{1(2)g},E_{1(2)g} \right)$& $2h$ \\
  $D_{2d}$ & $(E,E)$ & $2 d^2$ \\
  $D_{3d}$ & $\left(E_{g(u)},E_{g(u)}\right)$, $\left(E_{g},E_{u}\right)$ & $2 d^3$ \\
  $D_{4d}$ & $\left(E_{i},E_{i}\right)|_{i=1,2,3}$, $\left(E_{1},E_{3}\right)$ & $2 d^4$ \\
  $D_{6d}$ & $\left(E_{i},E_{i}\right)|_{i=1,2,3,4,5}$, $\left(E_{i},E_{6-i}\right)|_{i=1,2}$ & $2 d^6$ \\
  $C_{3v}$ & ($E,E$)  &  $2v^3$ \\
  $C_{4v}$ & ($E,E$)  &  $2 v^2$, $2 d^2$, $2 v^2 d^2$ \\
  $C_{6v}$ & $\left(E_{1(2)},E_{1(2)}\right)$ &   $2v^3 d^3$ \\
              & $ \left(E_{1},E_{2}\right)$  &  $2 v^3$, $2d^3$ \\
  $D_{3h}$ & $\left(E^{'('')},E^{'('')}\right)$ & $2 v^3$ \\ 
              & $\left(E^{'},E^{''}\right)$ & $2h$, $2 h v^3$ \\
  $D_{4h}$ & $\left(E_{g(u)},E_{g(u)}\right)$ & $2v^2$, $2 d^2$, $2 v^2 d^2$ \\
              & $\left(E_{g},E_{u}\right)$ &  $2h$, $2 h v^2$, $2 h d^2$, $2 h v^2 d^2$ \\
  $D_{6h}$ & $\left(E_{1g(u)},E_{1g(u)}\right)$, $\left(E_{2g(u)},E_{2g(u)}\right)$ &  $2 v^3 d^3$ \\
              & $\left(E_{1g(u)},E_{2g(u)}\right)$ & $2h v^3$, $2 h d^3$ \\
              &$\left(E_{1(2)g},E_{1(2)u}\right)$ & $2h$, $2 h v^3d^3$ \\
              & $\left(E_{1g(u)},E_{2u(g)}\right)$ & $2 v^3$, $2 d^3$ \\  
  $T_d$     & ($E,E$)  &  $2 d^6$ \\
              & $\left( T_1,T_2 \right)$  &  $3 d^6$ \\
  $T_h$     & $\left(E_g,E_u \right)$  &  $2 h^3$ \\
              & $\left( T_g,T_u \right)$  &  $3 h^3$ \\
  $O_h$     & $\left(E_g,E_g \right)$  &  $2 d^6$ \\
              & $\left( E_g,E_u \right)$  &  $2 h^3$, $2h^3 d^6$ \\
              & $\left( T_{1g(u)},T_{2g(u)} \right)$  &  $3 d^6$ \\
              & $\left( T_{1g(u)},T_{1g(u)} \right)$  &  $3 h^3 d^6$ \\
              & $\left( T_{1g(u)},T_{2u(g)} \right)$  &  $3 h^3$ \\
  \hline\hline
 \end{tabular}
\end{table*}
   
  Finally, we mention the cases that level schemes consist of different dimensional IRRs, such as $\left( \Gamma_{1a}, \Gamma_{2b}\right)$, $\left( \Gamma_{1a}, \Gamma_{3b}\right)$, and $\left( \Gamma_{2a}, \Gamma_{3b}\right)$. In this case, the above decomposition is not applicable because when we ignore a band splitting, a band always remains uncoupled with other bands, resulting in a metallic phase. Hence, we need to remove the unwanted energy bands away from the Fermi level by a band splitting. Then, we can engineer mirror-reflection symmetry-protected line nodes in a similar manner. 
    
\section{Topological numbers}
\label{app3}

A mirror-reflection symmetry-protected line node is attributed to the band degeneracy between the conduction and valence bands with opposite mirror-reflection eigenvalues. That is, on the $\sigma_{m}$-symmetric plane, the topological number can be given by counting  the number of occupied states with $\sigma_{m}=\lambda$ outside and inside of the nodal loops:
\begin{align}
\mathcal{Q}_{\lambda} = n_{{\rm occ},\lambda}^> - n_{{\rm occ},\lambda}^< \in \mathbb{Z}, 
\end{align}
where $ n_{{\rm occ},\lambda}^{>(<)}$ is the number of occupied states with $\sigma_{m}=\lambda$ outside (inside) a nodal loop. In the following, we consider time-reversal-invariant systems and show the $\mathbb{Z}_2$ topological number of line nodes, which is associated with  the $\mathbb{Z}_2$ topological number of TIs.

\subsection{\texorpdfstring{$\mathbb{Z}_2$}{Z2} topological number in the absence of SOI}
\label{app3-1}
From previous studies~\cite{kim-kane,yamakage16,chan16}, a DLN gives a nontrivial $\mathbb{Z}_2$ topological number in terms of the Berry phase, which links to the drumhead surface state and polarization (see {Appendix} \ref{toy}). The Berry phase in spinless systems is defined by~\cite{yamakage16}
\begin{align}
\Phi_{j} (k_l,k_i) = \oint_C d k_j \, {\rm tr} A_j (\bm{k}) - i \, {\rm tr } \ln B_{j} (k_l,k_i),  \label{eq:Berry}
\end{align}
where $A_j (\bm{k})$ and $ B_{j} (k_l)$ are the non-Abelian Berry connection and the sewing matrix defined by, respectively, 
\begin{align}
&[\bm{A}(\bm{k})]_{mn} = -i  \langle \bm{k},m|\partial_{\bm{k}}| \bm{k}, n \rangle, \\ 
&[B_{j} (k_l,k_i)]_{mn} = \langle \bm{k},m|B_j | \bm{k}-\bm{G}_j, n \rangle|_{k_j=\pi}.
\end{align}
Here, $i,j,l=1,2,3$, $|\bm{k}, n \rangle$ is the Bloch function with band index $n$, and the sewing matrix originates from a nontrivial periodic boundary condition:
\begin{align}
H(\bm{k}) = B_j^{\dagger} H(\bm{k}+\bm{G}_j)B_j,
\end{align}
instead of imposing the momentum dependence on a group operation. Using the Berry phase, the $\mathbb{Z}_2$ topological number is given by 
\begin{align}
 (-1)^{\nu(k_l, k_i)} = e^{i \Phi_{j}(k_l,k_i)},
 \label{nuli}
\end{align}
where $\nu(k_l, k_i)$ takes values of $0$ or $1$ due to the constraints from the spatial-inversion or mirror-reflection symmetry. 
If $\nu(k_l, k_i) =1$, a loop $C$ encircles a band degeneracy, 
implying that an odd number of DLNs penetrate into the inner side of $C$. 
In the following, 
we relate the $\mathbb{Z}_2$ topological numbers to spatial-inversion or mirror-reflection eigenvalues at a high symmetric momentum. 
Note that a similar argument was discussed in Refs.~\onlinecite{kim-kane,yamakage16}.
  For simplicity, we assume in the following that the nontrivial boundary condition occurs only for the $k_{\perp} $ direction, i.e., $H(k_{\perp},\bm{k}_{\parallel}) = B_{\perp}^{\dagger}H(k_{\perp}+G_{\perp},\bm{k}_{\parallel})B_{\perp}$, where $\bm{k}_{\parallel}$ is a momentum perpendicular to $k_{\perp}$.

\begin{figure}
\centering
\includegraphics[width=8cm]{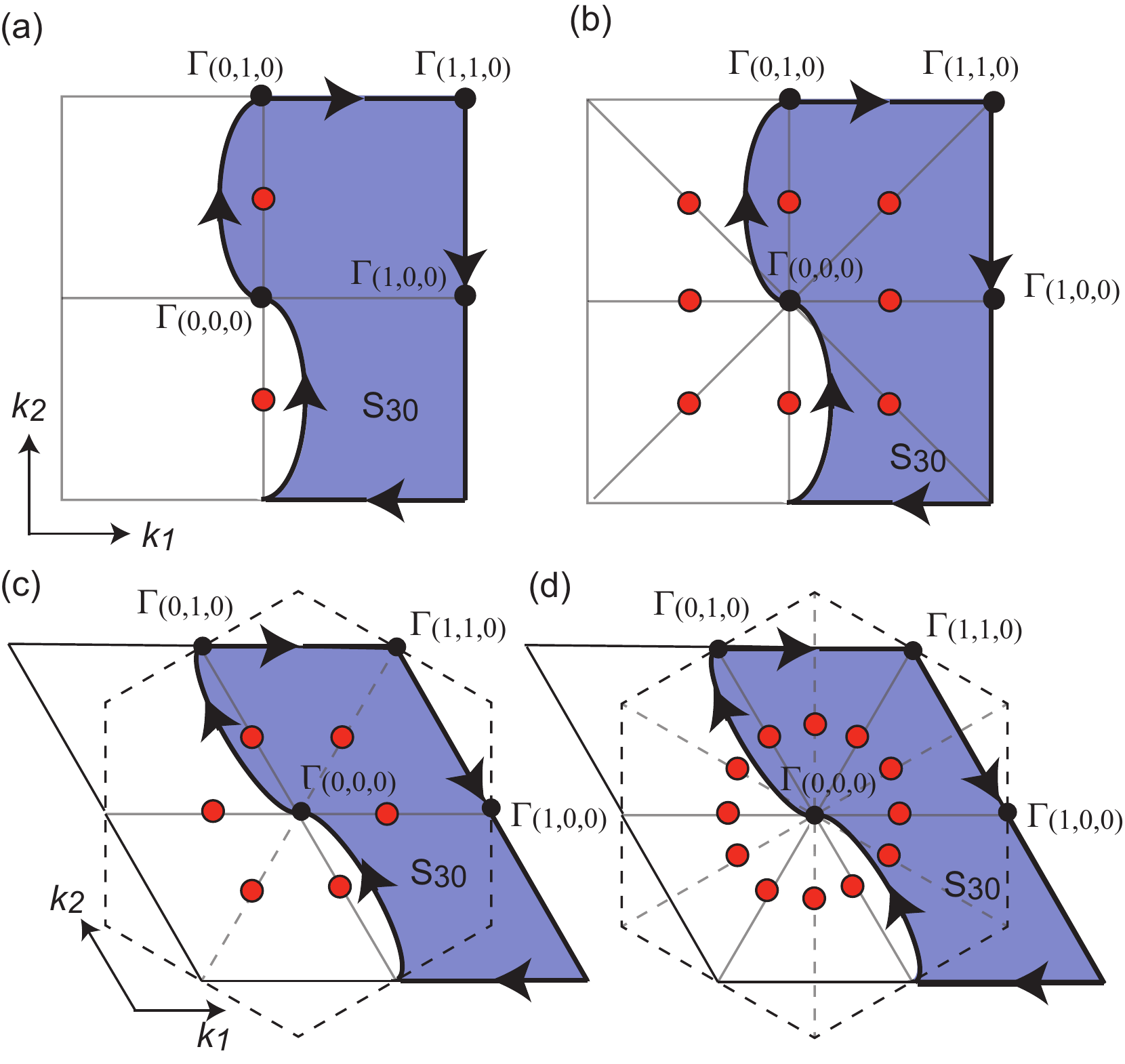}
\caption{The shape of loop $C= \partial S_{30}$ on the plane of $k_3 =0$, where the red dots indicate the position of DLNs. Here, we assume the crossing line nodes encircle $\bm{\Gamma}_{(0,0,0)}$ and change the path slightly in order to avoid the line node. Figure (a) represents the line node configuration $h$ of $C_s$, $C_{lh}$, $D_{lh}$, and $D_{2h}$ ($l=2,4,6$), (b) $v^2d^2$ of $D_{4h}$, (c) $d^3$ of $D_{3d}$, and (d) $v^3 d^3$ of $D_{6h}$. }
\label{fig:circle}
\end{figure}

First of all, consider centrosymmetric systems. The Hamiltonian hosts the spatial-inversion symmetry as
\begin{align}
 P H(\bm{k}) P^{\dagger} = H(-\bm{k}),
\end{align}
where $P$ is the spatial-inversion operator. Under the inversion operation, the non-Abelian Berry connection transforms as 
\begin{align}
 \bm{A} (\bm{k}) = -P(-\bm{k})^{\dagger}  \bm{A} (-\bm{k}) P(-\bm{k}) -i P(-\bm{k})^{\dagger} \partial_{\bm{k}} P(-\bm{k}),
\end{align}
where $[P(\bm{k})]_{nm} = \langle \bm{k},m|P \left| -\bm{k}, n \right\rangle$. As we consider the loop $C=\{(k_{\perp},\bm{\Gamma}_{\parallel})|-\pi \le k_{\perp} \le\pi \}$, where $\bm{\Gamma}_{\parallel}$ is a TRIM on the plane perpendicular to the $k_{\perp}$ direction, the integral of $\bm{A}(\bm{k})$ becomes
\begin{align}
&
 \int_{-\pi}^0 d k_{\perp} {\rm tr} A_{\perp} (k_{\perp},\bm{\Gamma}_{\parallel}) =-\int_{0}^{\pi} d k_{\perp} {\rm tr} A_{\perp} (k_{\perp},\bm{\Gamma}_{\parallel}) \notag \\
 &+ i \int_{0}^{\pi} dk_{\perp}{\rm tr} P(k_{\perp},\bm{\Gamma}_{\parallel})^{\dagger} \partial_{k_{\perp}} P(k_{\perp},\bm{\Gamma}_{\parallel}), 
\end{align}
which yields 
\begin{align}
 \int_{-\pi}^{\pi} d k_{\perp} {\rm tr} A_{\perp} (\bm{k}) = i \ln \frac{\det P(\pi, \bm{\Gamma}_{\parallel})}{\det P(0, \bm{\Gamma}_{\parallel})} + 2 \pi n , \ \ n \in \mathbb{Z}. \label{eq:intA}
\end{align}
Substituting Eq.~(\ref{eq:intA}) into Eq.(\ref{eq:Berry}), one obtains
\begin{align}
\Phi_{\perp} (\bm{\Gamma}_{\parallel}) = i \ln \frac{\det P(\pi, \bm{\Gamma}_{\parallel})'}{\det P(0, \bm{\Gamma}_{\parallel})}+ 2 \pi n , \ \ n \in \mathbb{Z},
\end{align}
where $ [P(\pi, \bm{\Gamma}_{\parallel})']_{mn} \equiv  [P(\pi, \bm{\Gamma}_{\parallel})B_{\perp} (\bm{\Gamma}_{\parallel})^{\dagger}]_{mn} = \langle (\pi,\bm{\Gamma}_{\parallel}),m|PB^{\dagger}| (\pi,\bm{\Gamma}_{\parallel}), n \rangle $. Therefore, when we choose the basis as $[P(\bm{\Gamma})]_{nm} = \xi_n(\bm{\Gamma}) \delta_{mn}$, this results in
\begin{align}
 (-1)^{\nu (\bm{\Gamma}_{\parallel})} = \prod_{n \in {\rm occ} } \xi_n(0,\bm{\Gamma}_{\parallel}) \xi_n(\pi ,\bm{\Gamma}_{\parallel}), \label{eq:parity}
\end{align}
which relates the $\mathbb{Z}_2$ topological number to the parity eigenvalues of the TRIMs. Here, $\xi_n(\bm{\Gamma})$ is the eigenvalue of $P$ at $\bm{\Gamma}$ and takes $\pm1$. For a surface $S_{i \eta} = \{(k_j,k_l,k_i=\eta\pi)| -\pi \le k_j,k_l \le \pi \}$ ($\eta=0,1$), the $\mathbb{Z}_2$ topological number of $C= \partial S_{i \eta}$ is given by 
\begin{align}
 (-1)^{N (S_{i \eta})} = \prod_{n_j,n_l=0,1 } \prod_{n \in {\rm occ} } \xi_n\left(\bm{\Gamma}_{\left(n_j,n_l, \eta \right)}\right),
\end{align}
where $N (S_{i \eta})$ is the number of DLNs penetrating into $S_{i \eta}$. Note that when a DLN crosses $\partial S_{i \eta}$, we slightly modify the path with spatial-inversion symmetry. (See Fig.~\ref{fig:circle} as an example.)

Next, consider noncentrosymmetric systems. In this case, we use the mirror-reflection symmetry instead of the spatial-inversion symmetry. The mirror-reflection symmetry satisfies
\begin{align}
 M H(k_{\perp},\bm{k}_{\parallel}) M^{\dagger} = H(-k_{\perp},\bm{k}_{\parallel}),
\end{align}
where $M$ is the mirror-reflection operator. Under the mirror-reflection operation, the non-Abelian Berry connection transforms as
\begin{align}
 \bm{A} (k_{\perp},\bm{k}_{\parallel}) = &-M(-k_{\perp},\bm{k}_{\parallel})^{\dagger}  \bm{A} (-k_{\perp},\bm{k}_{\parallel}) M(-k_{\perp},\bm{k}_{\parallel}) \notag \\ 
 &-i M(-k_{\perp},\bm{k}_{\parallel})^{\dagger} \partial_{\bm{k}} M(-k_{\perp},\bm{k}_{\parallel}),
\end{align}
where $[M(k_{\perp},\bm{k}_{\parallel})]_{nm} = \langle (k_{\perp},\bm{k}_{\parallel}),m|M|(-k_{\perp},\bm{k}_{\parallel}), n \rangle$. After integrating $\bm{A}(\bm{k})$ along $C=\{(k_{\perp},\bm{k}_{\parallel})|-\pi \le k_{\perp} \le\pi \}$ in a similar manner to the case of the spatial-inversion symmetry, it turns out that
\begin{align}
\Phi_{\perp} (\bm{k}_{\parallel}) = i \ln \frac{\det M(\pi, \bm{k}_{\parallel})'}{\det M(0, \bm{k}_{\parallel})}+ 2 \pi n , \ \ n \in \mathbb{Z},
\end{align}
where $ [ M(\pi, \bm{k}_{\parallel})']_{mn} \equiv  [M(\pi, \bm{k}_{\parallel})B_{\perp} (\bm{k}_{\parallel})^{\dagger}]_{mn} = \langle (\pi,\bm{k}_{\parallel}),m|PB^{\dagger}| (\pi,\bm{k}_{\parallel}), n \rangle $. When we choose the basis that satisfies  $[M(\Gamma_{\perp},\bm{k}_{\parallel})]_{nm} = \zeta_n(\Gamma_{\perp},\bm{k}_{\parallel}) \delta_{mn}$, where $\Gamma_{\perp}$ is a TRIM on a line perpendicular to the mirror-reflection symmetric plane and $ \zeta_n (\Gamma_{\perp},\bm{k}_{\parallel})=\pm1$, the $\mathbb{Z}_2$ topological number is described by the mirror-reflection eigenvalues:
\begin{align}
 (-1)^{\nu (\bm{k}_{\parallel})} = \prod_{n \in {\rm occ} } \zeta_n(0,\bm{k}_{\parallel}) \zeta_n(\pi ,\bm{k}_{\parallel}). \label{eq:mirror-parity}
\end{align}
Also, if we take a loop as $C= \partial S_{i \eta}$ which is the surface perpendicular to the mirror-reflection symmetric plane, one obtains
\begin{align}
 (-1)^{N (S_{i \eta})} = \prod_{n_j,n_l=0,1 } \prod_{n \in {\rm occ} } \zeta_n\left(\bm{\Gamma}_{\left(n_j,n_l, \eta \right)}\right). \label{eq:mirror-Z2}
\end{align}
In contrast with the spatial inversion cases, Eq.~(\ref{eq:mirror-Z2}) is applicable only if a DLN does not cross $\partial S_{i \eta}$. 

Finally, we mention the connection between the $\mathbb{Z}_2$ topological number and $\mathcal{Q}_{\lambda}$. When a Dirac nodal ring exists on the plane of $k_{\perp} =0$ and encircles a TRIM, it follows from the definition of $\mathcal{Q}_{\lambda}$ that
\begin{align}
 (-1)^{\mathcal{Q}_{\lambda}} & =  \prod_{n \in {\rm occ} } \zeta_n \left(0,\bm{\Gamma}_{\parallel, >}\right) \zeta_n \left(0 ,\bm{\Gamma}_{\parallel,<}\right) \notag  \\
  & =  \prod_{k_{\perp}=0,\pi }  \prod_{n \in {\rm occ} } \zeta_n(k_{\perp},\bm{\Gamma}_{\parallel, >}) \zeta_n(k_{\perp} ,\bm{\Gamma}_{\parallel,<}) \notag \\
 & =  (-1)^{\nu (\bm{\Gamma}_{\parallel, <}) },
 \label{Qnu}
\end{align} 
where $\bm{\Gamma}_{\parallel, > (<)}$ is a TRIM outside (inside) the nodal ring on the plane of $k_{\perp}=0$.

\subsection{\texorpdfstring{$\mathbb{Z}_2$}{Z2} topological number of topological insulators}
\label{app3-2}
Taking into account the SOI, some cases become topological insulators. Here, we prove the criterion of topological insulators, connecting the number of DLNs with the $\mathbb{Z}_2$ topological number of topological insulators. We start with the simplified expression:~\cite{yamakage16}
\begin{align}
 \nu_{i \eta} = \frac{\tilde{\Phi}_{jli\eta}(k_l=0)+\tilde{\Phi}_{jli\eta}(k_l=\pi)}{2 \pi} \, \mod 2, \label{eq:top-TI}
\end{align}
with
\begin{align}
\tilde{\Phi}_{jli\eta}(k_l) = \oint_C d k_j \, {\rm tr} A_j (\bm{k})|_{k_i = \eta \pi} 
- i \, {\rm tr } \ln B_{j} (k_l,\eta \pi),
\end{align}
where $\eta=0,1$ and $\tilde{\Phi}_{jli\eta}(k_l)$ includes the spin degrees of freedom, i.e., $\tilde{\Phi}_{jli\eta}(k_l) = 2 \Phi_{j}(k_i = \eta \pi, k_l)$ in the SOI-free limit. The $\mathbb{Z}_2$ topological number is obtained by $\nu_0 = \nu_{i0}+\nu_{i1} \mod 2$ and $\nu_{i} = \nu_{i 1}$. Note that Eq.~(\ref{eq:top-TI}) is applicable to noncentrosymmetric systems only when the $k_j$ axis is perpendicular to the mirror-reflection symmetric plane. 
When we choose a loop $C$ that does not cross DLNs in systems without SOI, the systems have a gap along $C$ with and without the SOI. Hence, the topological number does not change even when the SOI is turned off. Therefore, Eq.~(\ref{eq:top-TI}) is rewritten as
\begin{align}
 (-1)^{\nu_{i \eta} } =& \exp\left[i \left( \frac{\tilde{\Phi}_{jli\eta}(k_l=0)}{2} +\frac{\tilde{\Phi}_{jli\eta}(k_l=\pi)}{2}\right)\right] \notag \\
 = & \exp \left[i \left( \Phi_{j}(0,\eta \pi)+\Phi_{j}(\pi, \eta \pi) \right) \right] \notag \\
 = &(-1)^{ \nu (0,\eta \pi)+\nu (\pi,\eta \pi) } \notag \\
 = & (-1)^{N(S_{i \eta})}, \label{eq:DLN-TI}
\end{align}
where $N(S_{i \eta})$ represents the number of line nodes penetrating into the surface $S_{i \eta}=\{(k_j,k_l,k_i=\eta \pi)| -\pi \le k_j,k_l \le \pi \}$. Using the eigenvalues, 
Equation (\ref{eq:DLN-TI}) immediately leads to
\begin{align}
 \nu_{i \eta} = N(S_{i \eta}) \mod 2,
\end{align}
and Eqs. (1) and (2) in the main paper.
%%%%%%%%%%%%%%%%%%%%%%%%%%%
%%%%%%%%%%%%%%%%%%%%%%%%%%%
Equation~(\ref{eq:DLN-TI}) is described by, for centrosymmetric systems,
\begin{align}
  (-1)^{N(S_{i \eta})} = \prod_{k_j,k_l=0,\pi}\prod_{n \in {\rm occ} } \xi_n (k_j,k_l,k_i=\eta \pi),
\end{align}  
and, for noncentrosymmetric systems,
\begin{align}
  (-1)^{N(S_{i \eta})} = \prod_{k_j,k_l=0,\pi}\prod_{n \in {\rm occ} } \zeta_n (k_j,k_l,k_i=\eta \pi).
\end{align}  

Concretely, consider a crossing DLN encircling $\bm{\Gamma}_{(0,0,0)}$ in a centrosymmetric system. In this case, the $\mathbb{Z}_2$ topological number (\ref{eq:top-TI}) is calculated as $\nu_{11}=\nu_{21} = \nu_{31} =0$ and  $\nu_{10}=\nu_{20}=\nu_{30} = N(\bm{\Gamma}_{(0,0,0)}) \mod 2$, where $ N(\bm{\Gamma}_{(0,0,0)})$ is the number of line nodes encircling $\bm{\Gamma}_{(0,0,0)}$. Thus, one obtains
\begin{align}
 &\nu_1 = \nu_2 =\nu_3=0, \\
 &\nu_0 = N(\bm{\Gamma}_{(0,0,0)}) \mod 2.
 \end{align}
 Therefore, when the SOI makes a gap, the systems with an odd number of DLNs become topological insulators. 

\section{Drumhead surface states}
\label{toy}
The one-dimensional $\mathbb Z_2$ invariant Eq. (\ref{nuli}) partially guarantees the presence of drumhead surface states \cite{yamakage16, chan16}.
Here, we show an example of drumhead surface states for crossing-line-node semimetals.
%%%%%%%%%%%%%%%%%%%%%%%%%%%%%%%

\begin{figure}
\centering
\includegraphics{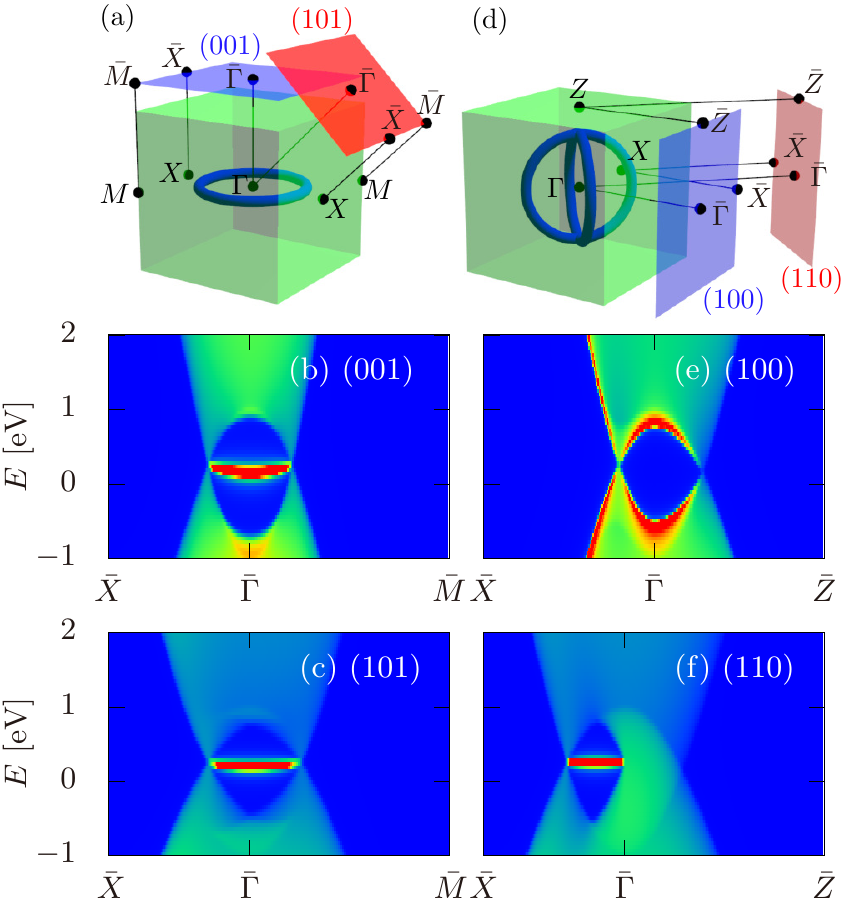}
\caption{
Line nodes and angle-resolved surface density of states for the $A_{1g}$--$A_{2u}$ [(a)--(c)] and $A_{1g}$--$B_{1g}$ [(d)--(f)] models. }
\label{eff}
\end{figure}

As an example, we examine two minimal models consisting of $A_{1g}$ and $A_{2u}$ orbitals ($A_{1g}$--$A_{2u}$ model) and of $A_{1g}$ and $B_{1g}$ orbitals ($A_{1g}$--$B_{1g}$ model) under the $D_{4h}$ point-group symmetry.
The Hamiltonians for these models are explicitly shown in the next section \ref{effective}.
%%%%%%%
Line nodes appear on the $k_z=0$ plane ($h$) in the former model [Fig. \ref{eff}(a)] while on the diagonal mirror planes ($k_x= \pm k_y$) but not on the vertical planes ($k_x, k_y = 0, \pi$) ($d^2$) in the latter model [Fig. \ref{eff}(b)].
The configurations, $h$ and $d^2$, of line nodes are consistent with the general theory discussed in the main manuscript.
Moreover, the general formulae Eqs. (\ref{eq:mirror-parity}) and (\ref{Qnu})  derived in the previous section tells us that the one-dimensional invariant $\nu_{[hkl]}(\bm k_\parallel)$, where the subscript $[hkl]$ denotes the direction of the integral path and $\bm k_\parallel$ is perpendicular to $[hkl]$, is obtained as follows: $\nu_{[001]}(\bm k_\parallel) = 1$ for the $A_{1g}$--$A_{2u}$ model and $\nu_{[110]}(\bm k_\parallel) = \nu_{[1 \bar 1 0]}(\bm k_\parallel) = 1$ for the $A_{1g}$--$B_{1g}$ model for $\bm k_\parallel$ located within the line nodes.
Additionally, in the latter model, $\nu_{[100]}(\bm k_{\parallel}) = \nu_{[010]}(\bm k_{\parallel}) = 0$ holds because there is no line node on the (100) and (010) planes.
Correspondingly, there exist surface states on the (001) surface of the former model and on the (110) and $(1 \bar 1 0)$ surfaces of the latter model while there is no surface state  on the (100) and (010) surfaces in the latter model, as numerically verified below.

%%%%%%%%%%%%%%%%%%%%%%%%%%%%%%%%%%%%%
%%%%%%%%%%%%%%%%%%%%%%%%%%%%%%%%%%%%%
We show the angle-resolved density of states on two different surfaces for the two models by calculating the surface Green's function \cite{miyata13, miyata15}. 
There exists a drumhead surface state within the line node on both the (001) [Fig. \ref{eff}(b)] and (101) [Fig. \ref{eff}(c)] surfaces of the $A_{1g}$--$A_{2u}$ model.
The $A_{1g}$--$B_{1g}$ model, on the other hand, has no surface state on the (100) surface, as shown in Fig. \ref{eff}(e), because the two line nodes have completely overlapped onto the (100) surface.
On the other surfaces, e.g., the (110) surface shown in Fig. \ref{eff}(f), surface states can emerge in the region in which the line nodes are not overlapping.
This result is also consistent with the general theory.

\section{Effective models}
\label{effective}

\subsection{\texorpdfstring{$A_{1g}$--$A_{2u}$}{A1g-A2u} model in \texorpdfstring{$D_{4h}$}{D4h}}

$k \cdot p$ Hamiltonian:
\begin{align}
	H(\bm k) &= c(\bm k) \sigma_0 + m(\bm k) \sigma_z + v k_z \sigma_y,
\end{align}
with
\begin{align}
	c(\bm k) &= c_0 + c_1 k_z^2 + c_2 (k_x^2+k_y^2),
	\\
	m(\bm k) &= m_0 + m_1 k_z^2 + m_2 (k_x^2+k_y^2),
\end{align}
up to the second order of the momentum $\bm k$.
The Hamiltonian is regularized on the cubic lattice into 
\begin{align}
	H'(\bm k) &= c'(\bm k) \sigma_0 + m'(\bm k) \sigma_z + v \sin k_z \sigma_y,
\end{align}
with
\begin{align}
	c'(\bm k) &= c_0 + 2 c_1 (1-\cos k_z) + 2 c_2 (2-\cos k_x -\cos k_y),
\end{align}
\begin{align}
	m'(\bm k) 
	&= m_0 + 2 m_1 (1-\cos k_z) 
	\nonumber\\ & \quad
	+ 2 m_2 (2 - \cos k_x - \cos k_y).
\end{align}

In order to calculate the surface electronic states, we set the semi-infinite Hamiltonian as
\begin{align}
 H(k_1, k_2)
 = \sum_n c^\dag_n \epsilon c_n 
 + \sum_n 
 \left( c^\dag_n t c_{n+1} + c^\dag_{n+1} t^\dag c_{n} \right).
\end{align}
On the (001) plane, for $k_1 = k_x$ and $k_2 = k_y$, the onsite and hopping matrices are given by
\begin{align}
 \epsilon_{001} &= 
 \left[
	 c_0 + 2 c_1 + 2 c_2 (2-\cos k_1 - \cos k_2)
 \right] \sigma_0
 \nonumber\\ & \quad
 +
 \left[
	 m_0 + 2 m_1 + 2 m_2 (2-\cos k_1 - \cos k_2)
 \right] \sigma_z,
 \\
 t_{001} &=
 -c_1 \sigma_0
 -m_1 \sigma_z
 - \mathrm i \frac{v}{2} \sigma_y.
\end{align}
On the (101) plane, for $k_1 = (k_x + k_z)/\sqrt 2$ and $k_2 = k_y$, we have
\begin{align}
 \epsilon_{101} &= 
 \left[
   c_0 + 2c_1 + 2c_2(2-\cos k_2)
 \right] \sigma_0
 \nonumber\\ & \quad
 +
 \left[
   m_0 + 2m_1 + 2m_2(2-\cos k_2)
 \right] \sigma_0,
\end{align}
and
\begin{align}
 t_{101}
 &=
 \left[
 	(-c_1-c_2) \cos \frac{k_1}{\sqrt 2} + \mathrm i (-c_1+c_2) \sin \frac{k_1}{\sqrt 2}
 \right] \sigma_0
 \nonumber\\ & 
 +
 \left[
 	(-m_1-m_2) \cos \frac{k_1}{\sqrt 2} + \mathrm i (-m_1+m_2) \sin \frac{k_1}{\sqrt 2}
 \right] \sigma_z
 \nonumber\\ & 
 +
 \left(
 	\frac{v}{2} \sin \frac{k_1}{\sqrt 2} - \mathrm i \frac{v}{2} \cos \frac{k_1}{\sqrt 2}
 \right) \sigma_y.
\end{align}
The parameters are set at $m_0 = -1$, $m_1=1.3$, $m_2=1.2$, $v=1.1$, $c_0=0$, $c_1=0.2$, $c_2 = 0.3$ in the calculation (Fig. \ref{eff}).

\subsection{\texorpdfstring{$A_{1g}$--$B_{1g}$}{A1g-B1g} model in \texorpdfstring{$D_{4h}$}{D4h}}

\begin{align}
  H(\bm k) = c(\bm k) \sigma_0 + m(\bm k) \sigma_z + v \left(k_x^2 - k_y^2 \right) \sigma_x,
\end{align}

\begin{align}
 H'(\bm k) = c'(\bm k) \sigma_0 + m'(\bm k) \sigma_z + 2 v \left(
  -\cos k_x + \cos k_y
 \right) \sigma_x,
\end{align}

\begin{align}
 \epsilon_{100} &= 
 \left[
 c_0 + 2 c_1 (1 - \cos k_z) + 2 c_2 (2 -\cos k_y)
 \right]
 \sigma_0
 \nonumber\\ & \quad
 +
  \left[
 m_0 + 2 m_1 (1 - \cos k_z) + 2 m_2 (2 -\cos k_y)
 \right]
 \sigma_z
 \nonumber\\ & \quad
 +
 2 v \cos k_y \sigma_x,
\end{align}

\begin{align}
 t_{100} = 
 -c_2 \sigma_0 - m_2 \sigma_z - v \sigma_x,
\end{align}

\begin{align}
 \epsilon_{1  1 0} &= \left[
   c_0 + 2 c_1 (1-\cos k_z)
 \right] \sigma_0
 \nonumber\\ & \quad
 +
 \left[
   m_0 + 2 m_1 (1-\cos k_z)
 \right] \sigma_z,
\end{align}

\begin{align}
 t_{110}
 = -2 c_2 \cos\frac{k_1}{\sqrt 2} \sigma_0 
   -2 m_2 \cos\frac{k_1}{\sqrt 2} \sigma_z
   +\mathrm i 2 v \sin \frac{k_1}{\sqrt 2} \sigma_x,
\end{align}

\section{Rare-earth trihydrides}
\label{RH3}

HoD$_3$-structured materials without correlations ubiquitously exhibit crossing line nodes in the band gap. 
We show the energy band structure of LuH$_3$, which has 14 $f$ electrons, with the HoD$_3$ structure as another example of a crossing-line-node semimetal.
The lattice constant is taken from the calculated value in Ref. \onlinecite{kong12}.
\begin{figure}
\centering
\includegraphics{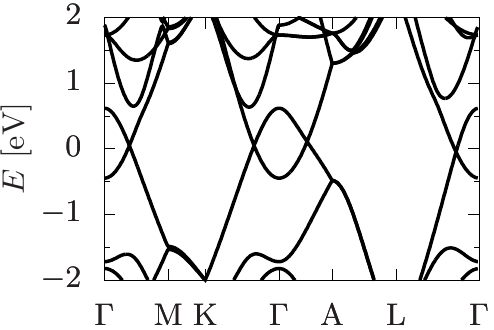}
\caption{Energy band of  LuH$_3$ with the $P \bar 3 c 1$ symmetry.}
\label{LuH3}
\end{figure}
The obtained first-principles band structure shown in Fig. \ref{LuH3} is quite similar to that for YH$_3$ (Fig. \ref{dos}) without correlation effects, i.e., three crossing line nodes ($d^3$ of $D_{3d}$ in Table \ref{tab:PG-LN}) are realized.
%%%%%%%%%%%%%%%%%%%%%%%%%%%%%%%
\begin{figure}
\centering
\includegraphics{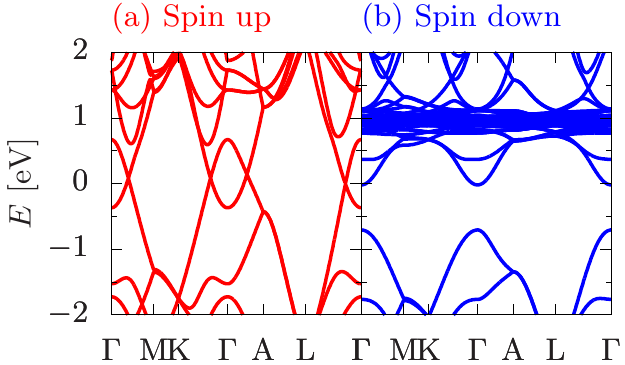}
\caption{Energy bands for (a) spin up and (b) spin down states in ferromagnetic GdH$_3$.}
\label{GdH3}
\end{figure}
%%%%%%%%%%%%%%%%%%%%%%%%%%%%%%%
One more example is ferromagnetic GdH$_3$ with the HoD$_3$ structure, where Gd's have $S=7/2$ spins.
The energy bands for spin up (majority) and for spin down (minority) are shown in Fig. \ref{GdH3}. 
The $f$ electrons migrate from the Fermi level to higher-energy regions.
The remaining spin-up state hosts three crossing line nodes, as with LuH$_3$ (Fig. \ref{LuH3}), while the spin-down state is insulating.
The resulting state is a crossing-line-node ($d^3$) half semimetal.
Note that, in the actual material of GdH$_3$, the antiferromagnetic state is more stable \cite{kong13} than the ferromagnetic state as the ground state and has been observed below $T_{\rm N} = 1.8$\,K \cite{flood77, carlin80}.

\bibliography{yh3}

\clearpage
\end{document}